\documentclass[structabstract]{aa}
\usepackage{graphicx}
\usepackage{txfonts}
\usepackage{natbib}
\usepackage{comment}

\newcommand{\maa}{\alpha \alpha} 
\newcommand{\mga}{\gamma \alpha}

\def\Msolar{$M_{\odot}$}

\newcommand{\perpccubed}{\,{\rm pc^{-3}}}
\newcommand{\Mo}{M_{\odot}}
\newcommand{\Ro}{R_{\odot}}
\newcommand{\thickhline}{\noalign{\hrule height 0.8pt}}

\begin{document}
\title{The effect of common-envelope evolution on the visible population of post-common-envelope binaries}
\subtitle{}
\titlerunning{The effect of CE evolution on the visible population of PCEBs}

\author{S.~Toonen \inst{1} \and G.~Nelemans \inst{1,2}}
\institute{Department of Astrophysics, IMAPP, Radboud University Nijmegen, PO Box 9010, 6500 GL Nijmegen, The Netherlands \\
    \email{silviato@astro.ru.nl}
	\and Instituut voor Sterrenkunde, KU Leuven, Celestijnenstraat 200D, 3001 Leuven, Belgium}
\date{Received 22 April 2013; Accepted 1 July 2013}
\abstract
{An important ingredient in binary evolution is the common-envelope (CE) phase. Although this phase is believed to be responsible for the formation of many close binaries, the process is not well understood.}
{We investigate the characteristics of the population of post-common-envelope binaries (PCEB). As the evolution of these binaries and their stellar components are relatively simple, this population can be directly used to constraint CE evolution.}
{We use the binary population synthesis code SeBa to simulate the current-day population of PCEBs in the Galaxy. We incorporate the selection effects in our model that are inherent to the general PCEB population and that are specific to the SDSS survey, which enables a direct comparison for the first time between the synthetic and observed population of visible PCEBs. }
{We find that selection effects do not play a significant role on the period distribution of visible PCEBs. To explain the observed dearth of long-period systems, the $\alpha$-CE efficiency of the main evolutionary channel must be low. In the main channel, the CE is initiated by a red giant as it fills its Roche lobe in a dynamically unstable way. 
Other evolutionary paths cannot be constrained more. Additionally our model reproduces well the observed space density, the fraction of visible PCEBs amongst white dwarf (WD)-main sequence (MS) binaries, and the WD mass versus MS mass distribution, but overestimates the fraction of PCEBs with helium WD companions.} 
{}

\maketitle

\section{Introduction}
\label{sec_ch4:intro}
Many close binaries are believed to have encountered an unstable phase of mass transfer leading to a common-envelope (CE) phase \citep{Pac76}. The CE phase is a short-lived phase in which the envelope of the donor star engulfs the companion star. Subsequently, the companion and the core of the donor star spiral inward through the envelope. If sufficient energy and angular momentum is transferred to the envelope, it can be expelled, and the spiral-in phase can be halted before the companion merges with the core of the donor star. The CE phase plays an essential role in binary star evolution and, in particular, in the formation of short-period systems that contain compact objects, such as post-common-envelope binaries (PCEBs), cataclysmic variables (CVs), the progenitors of Type Ia supernovae, and gravitational wave sources, such as double white dwarfs. 

Despite of the importance of the CE phase and the enormous efforts of the community, all effort so far have not been successful in understanding the phenomenon in detail. 
Much of the uncertainty in the CE phase comes from the discussion of which and how efficient certain energy sources can be used to expel the envelope \citep[e.g. orbital energy and recombination energy,][]{Ibe93, Han95, Web08}, or if angular momentum can be used \citep{Nel00, Van06}. 
Even though hydrodynamical simulations of parts of the CE phase \citep{Ric08,Ric12,Pas12} have become possible, simulations of the full CE phase are not feasible yet due to the wide range in time and length scales that are involved \citep[see][for reviews]{Taa00, Taa10, Iva13}.

In this study, a binary population synthesis (BPS) approach is used to study CE evolution in a statistical way. BPS is an effective tool to study mechanisms that govern the formation and evolution of binary systems and the effect of a mechanism on a binary population. 
Particularly interesting for CE research is the population of PCEBs (defined here as close, detached WDMS-binaries with periods of less than 100d that underwent a CE phase) for which the evolution of the binary and its stellar components is relatively simple. Much effort has been devoted to increase the observational sample and to create a homogeneously selected sample of PCEBs \citep[e.g.][]{Sch03, Reb07,Neb11}. 

In recent years, it has become clear that there is a discrepancy between PCEB observations and BPS results. BPS studies \citep{deK93, Wil04, Pol07, Dav10} predict the existence of a population of long period PCEBs ($>$10d) that have not been observed \citep[e.g.][]{Neb11}. It is unclear if the discrepancy is caused by a lack of understanding of binary formation and evolution or by observational biases.  
This study aims to clarify this by considering the observational selection effects that are inherent to the PCEB sample into the BPS study. Using the BPS code SeBa, a population of binary stars is simulated with a realistic model of the Galaxy and magnitudes and colors of the stellar components. 
In sect.\,\ref{sec_ch4:met}, we describe the BPS models, and in sect.\,\ref{sec_ch4:res}, we present the synthetic PCEB populations generated by the models. In sect. \,\ref{sec_ch4:res_sdss}, we incorporate the selection effects in our models that are specific to the population of PCEBs found by the SDSS. Comparing this to the observed PCEB sample \citep{Neb11, Zor11b} leads to a constrain on CE evolution, which will be discussed in sect.\,\ref{sec_ch4:concl}.

\section{Method}
\label{sec_ch4:met}
 
\subsection{SeBa  - a fast stellar and binary evolution code}
\label{sec_ch4:seba}

We employ the binary population synthesis code SeBa \citep{Por96, Nel01, Too12} to simulate a large amount of binaries. We use SeBa to evolve stars from the zero-age main sequence until remnant formation. At every timestep, processes as stellar winds, mass transfer, angular momentum loss, common envelope, magnetic braking, and gravitational radiation are considered with appropriate recipes. Magnetic braking \citep{Ver81} is based on \citet{Rap83}. A number of updates to the code has been made since \citet{Too12}, which are described in Appendix\,\ref{sec_ch4:ap_seba}. The most important update concerns the tidal instability \citep{Dar1879, Hut80} in which a star is unable to extract sufficient angular momentum from the orbit to remain in synchronized rotation, leading to orbital decay and a CE phase. 
Instead of checking at RLOF, we assume that a tidal instability leads to a CE phase instantaneously when tidal forces become affective i.e. when the stellar radius is less than one-fifth of the periastron distance.

SeBa is incorporated in the Astrophysics Multipurpose Software
Environment, or AMUSE. This is a component library with a homogeneous
interface structure and can be downloaded for free at {\tt
amusecode.org} \citep{Por09}.

\subsection{The initial stellar population}
The initial stellar population is generated on a Monte Carlo based approach, according to appropriate distribution functions. These are
\begin{equation}
\begin{array}{lccl}
\rm Prob(M_i) =        & \rm KTG93	 	& \rm for & 0.95\Mo \leqslant M_i \leqslant 10\Mo, \\ 
\rm Prob(q_i) \propto  & \rm const 		& \rm for & 0<q_i \leqslant 1,\\ 
\rm Prob(a_i) \propto  & a_{i}^{-1}\	\rm (A83)	& \rm for & 0\leqslant \rm log\ a_i/\Ro \leqslant 6,\\
\rm Prob(e_i) \propto  & 2e_i\ \rm (H75)			& \rm for & 0 \leqslant e_i \leq 1,\\
\end{array}
\label{eq_ch4:init}
\end{equation}
where $M_i$ is the initial mass of the more massive star in a specific binary system, the initial mass ratio is defined as $q_i\equiv m_i/M_i$ with $m_i$ the initial mass of the less massive star, $a_i$ is the initial orbital separation and $e_i$ the initial eccentricity. Furthermore, KTG93 represents \citet{Kro93}, A83 \citet{Abt83}, and H75 \citet{Heg75}.

\subsection{Common-envelope evolution}

For CE evolution, two evolutionary models are adopted that differ in their treatment of the CE phase. The two models are based on a combination of different formalisms for the CE phase. The $\alpha$-formalism \citep{Tut79} is based on the energy budget, whereas the $\gamma$-formalism \citep{Nel00} is based on the angular momentum balance. 
 In model $\alpha\alpha$, the $\alpha$-formalism is used to determine the outcome of the CE phase. For model $\mga$, the $\gamma$-prescription is applied unless the CE is triggered by a tidal instability rather than dynamically unstable Roche lobe overflow \citep[see][]{Too12}.  

In the $\alpha$-formalism, the $\alpha$-parameter describes the efficiency with which orbital energy is consumed to unbind the CE according to
\begin{equation}
E_{\rm gr} = \alpha (E_{\rm orb,init}-E_{\rm orb,final}),
\label{eq_ch4:alpha-ce}
\end{equation}
where $E_{\rm orb}$ is the orbital energy and $E_{\rm gr}$ is the binding energy of the envelope.  The orbital and binding energy are as shown in \citet{Web84}, where $E_{\rm gr}$ is approximated by
\begin{equation}
E_{\rm gr} = \frac{GM_{\rm d} M_{\rm d,env}}{\lambda R},
\label{eq_ch4:Egr}
\end{equation} 
where $M_{\rm d}$ is the donor mass, $M_{\rm d, env}$ is the envelope mass of the donor star, $R$ is the radius of the donor star, and in principle, $\lambda$ depends on the structure of the donor \citep{deK87, Dew00, Xu10, Lov11}. 

\begin{table}
\caption{Common-envelope prescription and efficiencies for each model.}
\begin{tabular}{lcc}
\thickhline
 & $\gamma$ & $\alpha\lambda$\\
Model $\mga$1 & 1.75 & 2 \\
Model $\maa$1 & - & 2 \\
Model $\mga$2 & 1.75 & 0.25 \\
Model $\maa$2& - & 0.25 \\
\thickhline
\hline
\end{tabular}
\label{tbl_ch4:models}
\end{table}

In the $\gamma$-formalism, $\gamma$-parameter describes the efficiency with which orbital angular momentum is used to expel the CE according to
\begin{equation}
\frac{J_{\rm b, init}-J_{\rm b,  final}}{J_{\rm b,init}} = \gamma \frac{\Delta M_{\rm d}}{M_{\rm d}+ M_{\rm a}},
\end{equation} 
where $J_{\rm b,init}$ and $J_{\rm b,final}$ are the orbital angular momentum of the pre- and post-mass transfer binary respectively, and $M_{\rm a}$ is the mass of the companion. 

The motivation for the $\gamma$-formalism comes from the observed distribution of double WD systems that could not be explained by the $\alpha$-formalism nor stable mass transfer for a Hertzsprung gap donor star \citep[see][]{Nel00}. The idea is that angular momentum can be used for the expulsion of the envelope, when there is a large amount of angular momentum available such as in binaries with similar-mass objects. However, the physical mechanism remains unclear. 

In the standard model in SeBa, we assume $\gamma = 1.75$ and $\alpha\lambda=2$, based on the evolution of double WDs \citep{Nel00, Nel01}. However lower CE efficiencies have been claimed \citep{Zor10}, and therefore, we construct a second set of models assuming $\alpha\lambda=0.25$. See Table\,\ref{tbl_ch4:models} for an overview of the models that are used in this paper. 

\subsection{Galactic model}
\label{sec_ch4:gal}
When studying populations of stars that are several Gyr old on average, the star formation history of the Galaxy becomes important. 
We follow \citet{Nel04} in taking a realistic model of the Galaxy based on \citet{Boi99}. 
In this model, the star formation rate is a function of time and position in the Galaxy. It peaks early in the history of the Galaxy and has decreased substantially since then. 
We assume the Galactic scale height of our binary systems to be 300 pc \citep{Roe07b, Roe07a}. The resulting population of PCEBs at a time of 13.5 Gyr is analysed.

\subsection{Magnitudes}
For WDs, the absolute magnitudes are taken from the WD cooling curves of pure hydrogen atmosphere models \citep[][and references therein\footnote{See also http://www.astro.umontreal.ca/$\sim$bergeron/CoolingModels.}]{Hol06, Kow06, Tre11}. These models cover the range of temperatures of $T_{\rm eff} =1500-100000$K and of surface gravities of $log\ =7.0-9.0$ for WD masses between 0.2 and 1.2\Msolar. 
For MS stars of spectral type A0-M9, we adopt the absolute magnitudes as given by \citet{Kra07}. Overall, the colours correspond well to colours from other spectra, such as the observational spectra from \citet[][with colors by \citet{Cov07}]{Pic98} and synthetic spectra \citep{Mun05} from Kurucz's code \citep{Kur81, Kur93}. 
For both the MS stars and WDs, we linearly interpolate between the brightness models. For MSs and WDs that are not included in the grids, the closest gridpoint is taken. 

To convert absolute magnitudes to apparent magnitudes, the distance from the sun is used as given by the Galactic model. Furthermore, we adopt the total extinction in the V filter band from \citet{Nel04}, which is based on Sandage's extinction law \citep{San72}. We assume the Galactic scale height of the dust to be 120 pc  \citep{Jon11}. 
To evaluate the magnitude of extinction in the different bands of the ugriz-photometric system, we use the conversion of \citet{Sch98}, which are based on the extinction laws of \citet{Car89} and \citet{ODo94} with $R_V=3.1$.

\subsection{Selection effects}
\label{sec_ch4:vis}
We assume that WDMS binaries can be observed in the magnitude range 15-20 in the g-band. 
As WDs are inherently blue and MS are inherently red, we assume that WDMS binaries can be distinguished from single MS stars if
\begin{equation}
\Delta g \equiv g_{\rm WD} - g_{\rm MS} < 1,
\label{eq_ch4:dg}
\end{equation}
where $g_{\rm WD}$ and $g_{\rm MS}$ are the magnitude in the g band of the WD and the MS respectively, and distinguished from single WDs if
\begin{equation}
\Delta z \equiv z_{\rm WD} - z_{\rm MS} > -1,
\label{eq_ch4:dz}
\end{equation}
where $z_{\rm WD}$ and $z_{\rm MS}$ are the z band magnitudes of the WD and the MS respectively. 
The g-band is used instead of the u-band, because the u-g colours of late-type MS stars are fairly uncertain \citep{Mun05, Boc07}. The effect of (varying) the cuts will be discussed in forthcoming sections.

\section{Results} 
\label{sec_ch4:res}

    \begin{figure*}
    \centering
    \begin{tabular}{ccc}
	\includegraphics[width=0.3\textwidth]{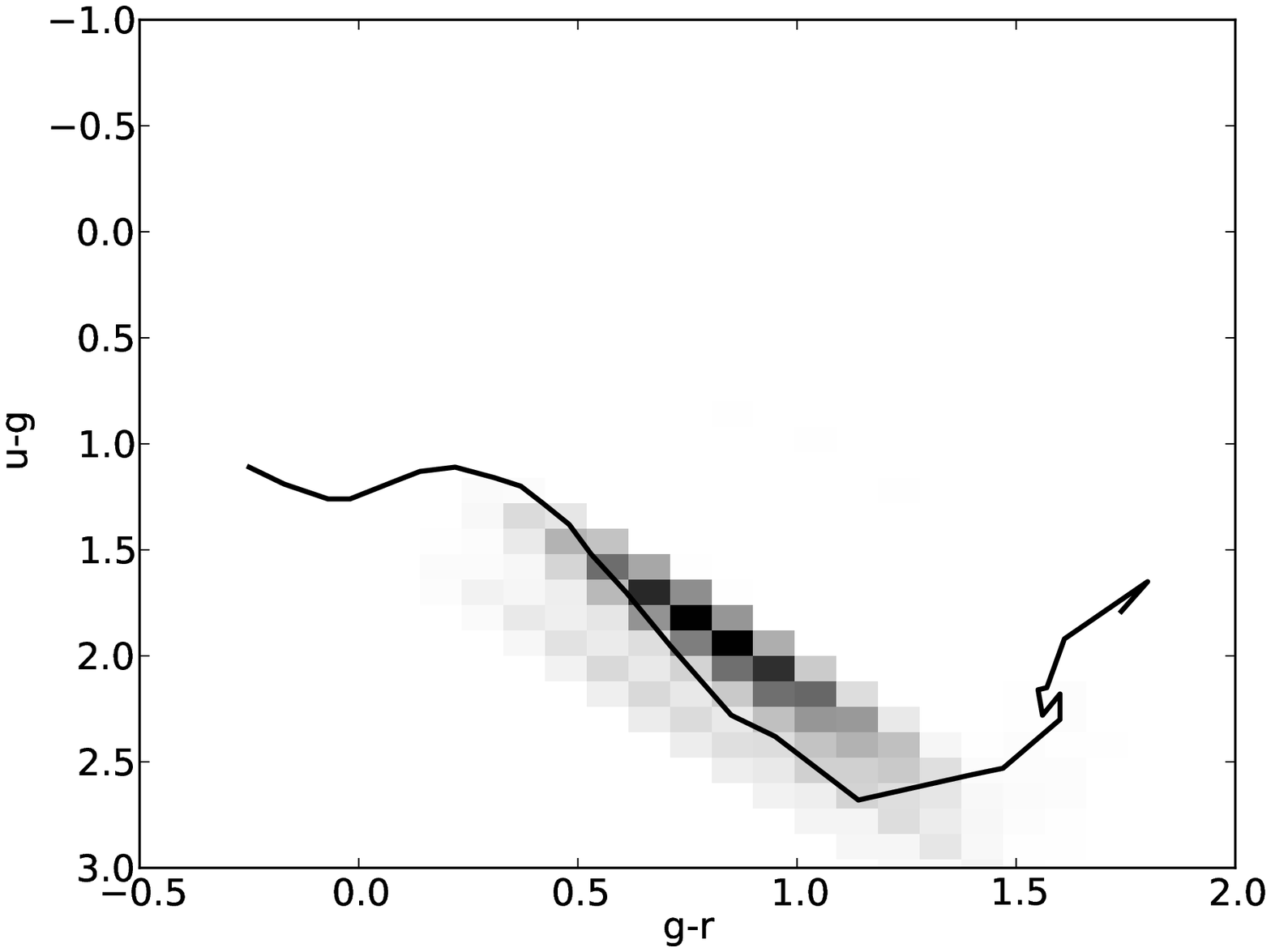} &
	\includegraphics[width=0.3\textwidth]{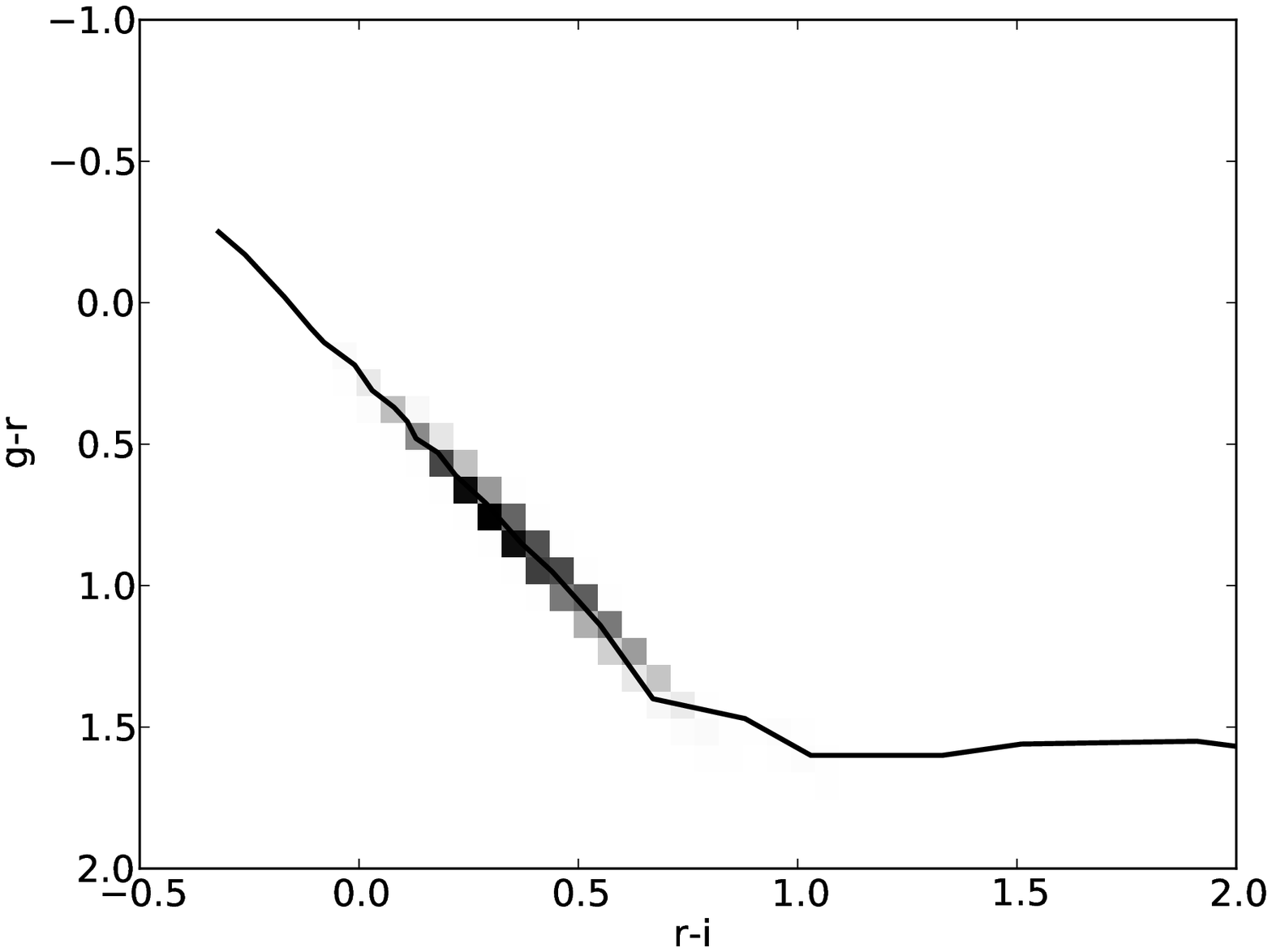} & 
	\includegraphics[width=0.3\textwidth]{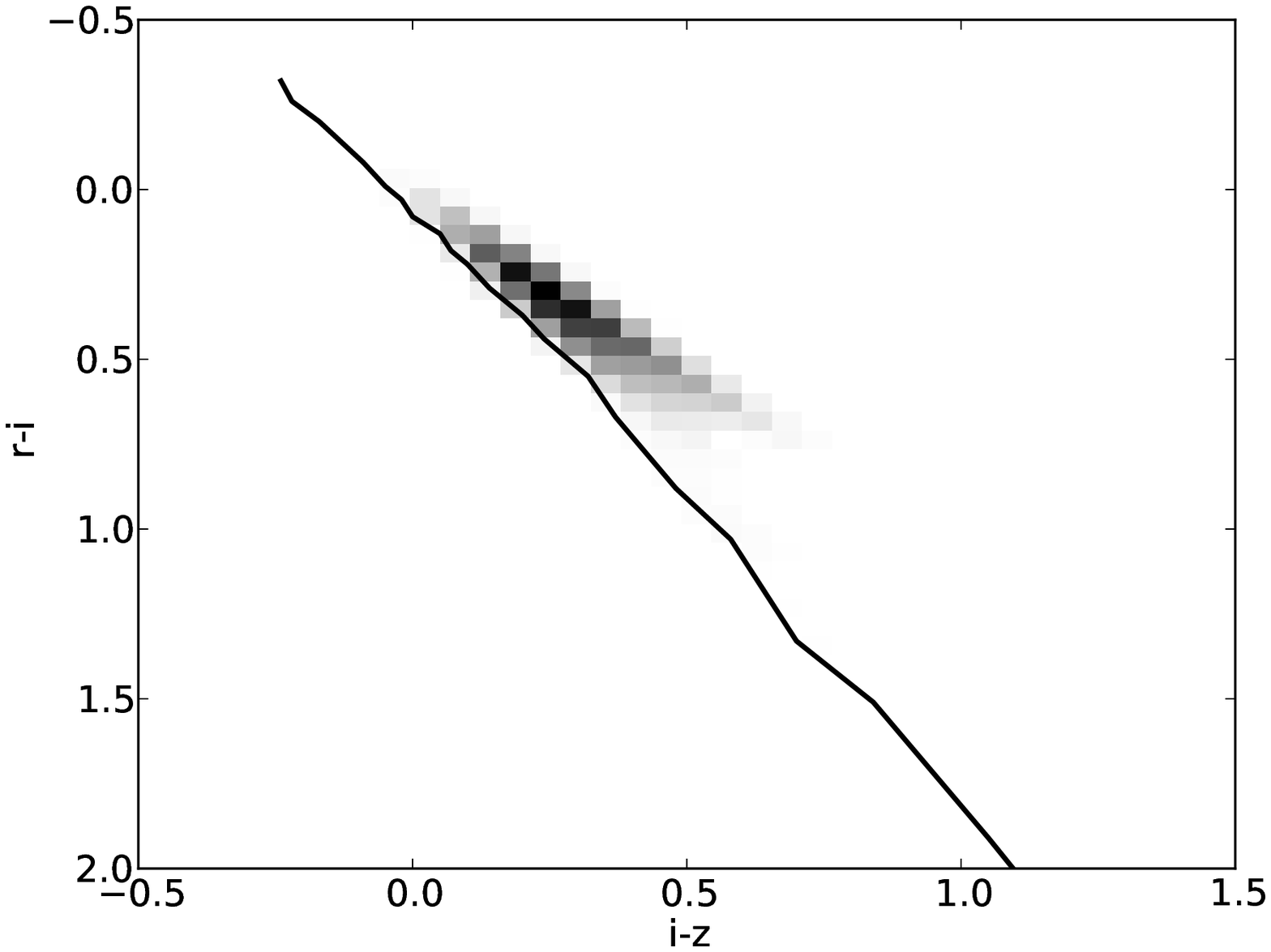} \\
	\end{tabular}
    \caption{Color-color diagrams for the full population of PCEBs with orbital periods less than 100d and for a limiting magnitude of $g=15-20$ for model $\maa$2. On the left, it shows the u-g vs. g-r diagram, in the middle, the g-r vs. r-i diagram, and on the right, the r-i vs. i-z diagram. The intensity of the grey scale corresponds to the density of objects on a linear scale. The solid line corresponds to the unreddened MS from A-type to M-type MS stars.} 
    \label{fig_ch4:color_color_all}
    \end{figure*}

    \begin{figure*}
    \centering
    \begin{tabular}{ccc}
	\includegraphics[width=0.3\textwidth]{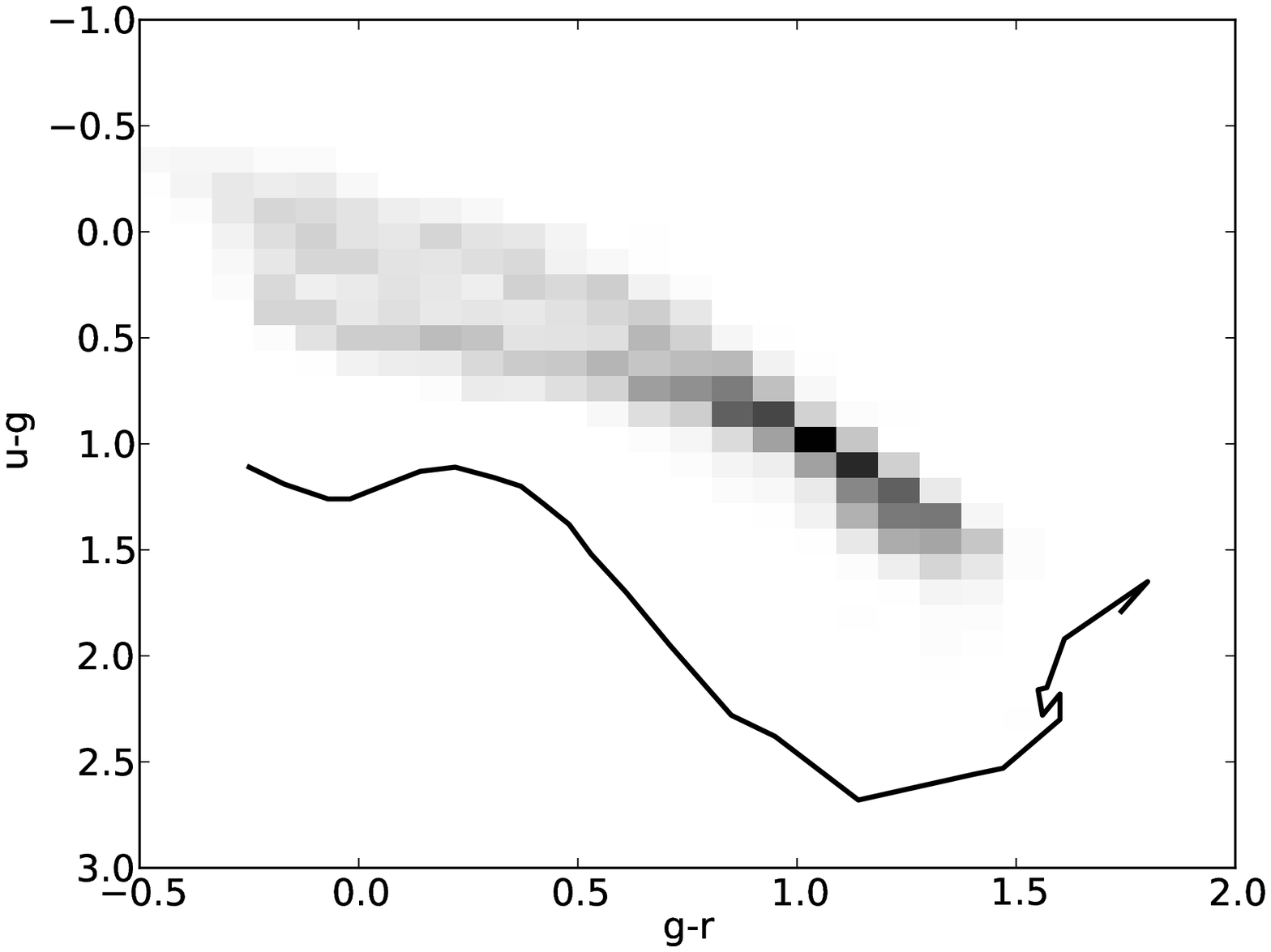} &
	\includegraphics[width=0.3\textwidth]{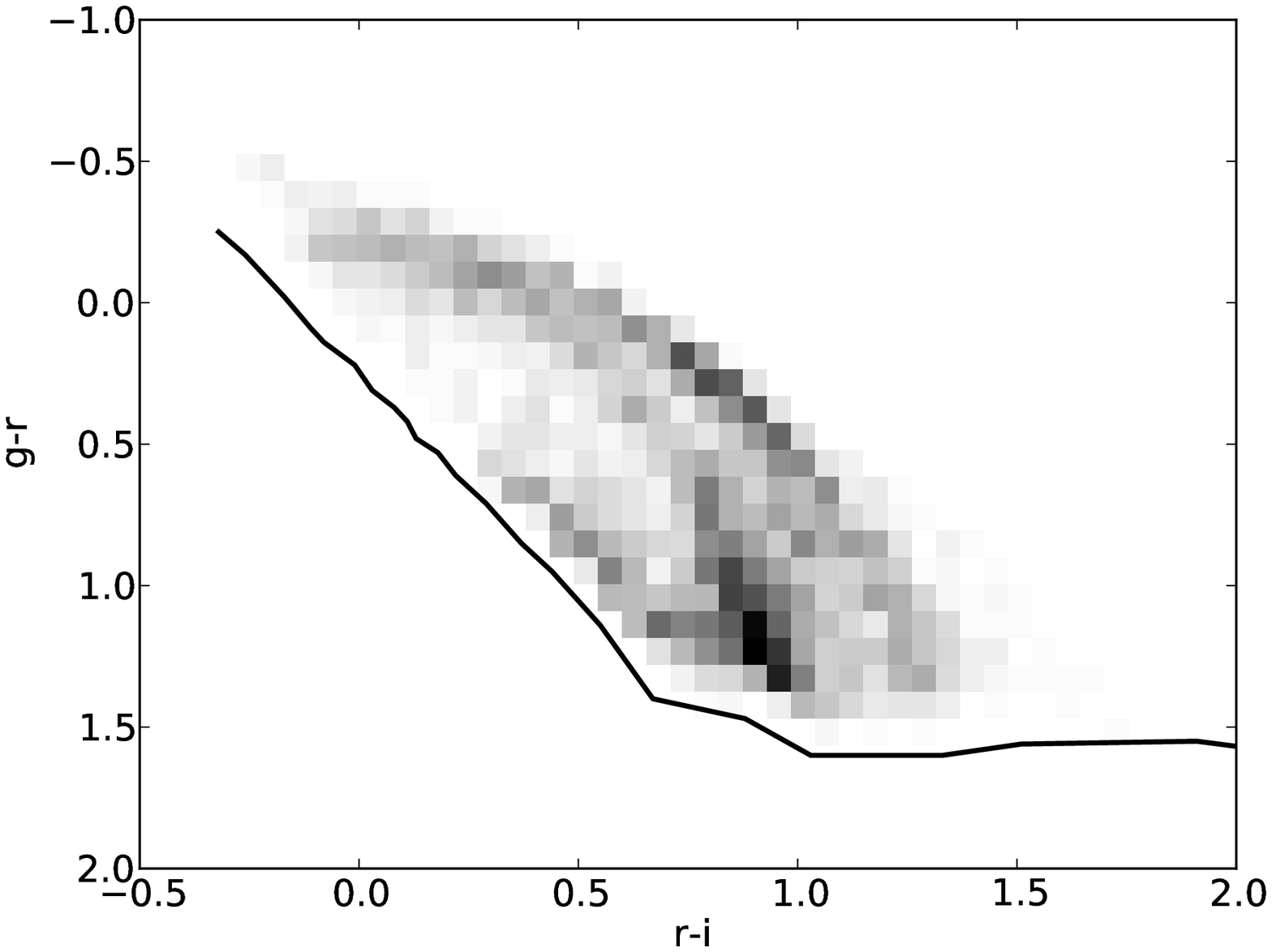} & 
	\includegraphics[width=0.3\textwidth]{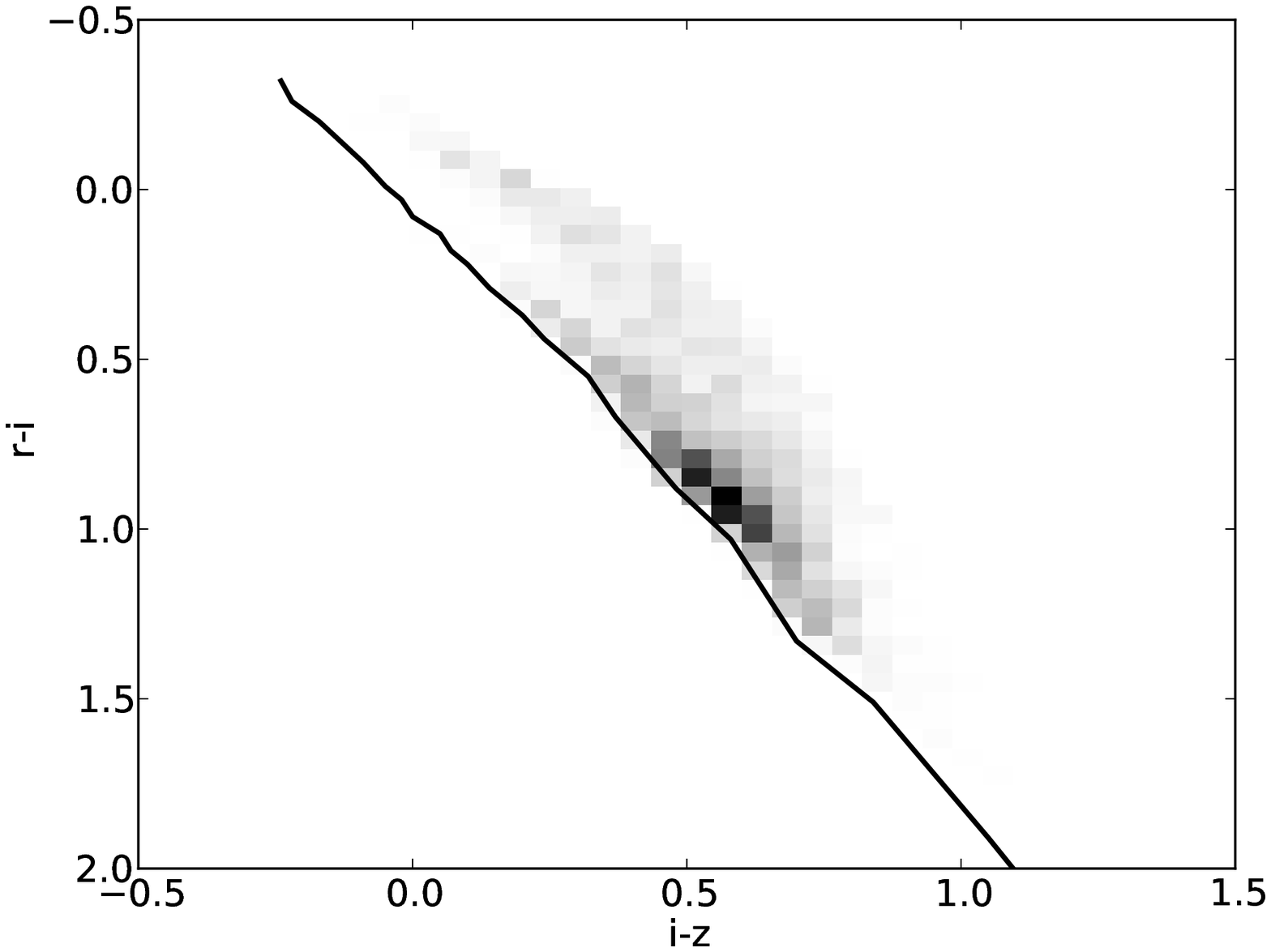} \\
	\end{tabular}
    \caption{Color-color diagrams for the visible population of PCEBs for model $\maa$2. On the left it shows the u-g vs. g-r diagram, in the middle the g-r vs. r-i diagram and on the right the r-i vs. i-z diagram. The intensity of the grey scale corresponds to the density of objects on a linear scale. The solid line corresponds to the unreddened MS from A-type to M-type MS stars.} 
    \label{fig_ch4:color_color}
    \end{figure*}

Figures\,\ref{fig_ch4:color_color_all}~and~\ref{fig_ch4:color_color} show the full and visible population of PCEBs in ugriz color-color space for model $\maa$2. 
The full population of PCEBs lies close to the unreddened MS. Most PCEB systems will be observed as apparent single MS stars. 
On the other hand, the visible population of PCEBs is by construction clearly distinguished from the MS in the u-g vs. g-r color-color diagram. In the r-i vs. i-z diagram and g-r vs. r-i diagram, most visible systems lie close to the MS indicating that the WD components are generally cold \citep[see][Fig\,2]{Aug08}.
The u-g vs. g-r diagram also shows that the majority of systems is relatively red confirming that samples of PCEBs that are discovered by their blue colors \citep[e.g.][]{Sch03}, are severely biased and incomplete. 
The color-color diagrams for model $\maa$1, model $\mga$1, and model $\mga$2 are very comparable to those of model $\maa$2. 

The space density of visible PCEBs follows directly from our models where the position of the PCEBs in the Galaxy is given by the Galactic model (see sect.\,\ref{sec_ch4:gal}). The space density (see Table\,\ref{tbl_ch4:space_density}) is calculated in a cylindrical volume with height above the plane of 200pc and radii of 200pc and 500pc centred on the Sun. At small distances ($\lesssim 100$pc) from the Sun, our data is noisy due to low number statistics, and at larger distances, the PCEB population is magnitude limited. The observed space density of PCEBs $(6-30)\cdot 10^{-6}\perpccubed$ \citep{Sch03} is fairly uncertain and consistent with all BPS models.

\begin{table}
\caption{The space density of visible PCEBs within 200 and 500pc from the Sun in $10^{-6}\perpccubed$for different models of CE evolution. }
\begin{tabular}{lcc}
\thickhline
& within 200pc & within 500 pc\\
Model $\mga$1 & 13 & 9.0\\
Model $\maa$1 & 15 & 12\\
Model $\mga$2 & 5.8 & 5.2 \\
Model $\maa$2 & 4.9 & 4.0 \\
\\
Observed & \multicolumn{2}{c}{6-30$^1$}\\
\thickhline
\hline
\end{tabular}
\label{tbl_ch4:space_density}
\tablefoot{$^1$ \citet{Sch03}.}
\end{table}

\begin{figure*}
    \centering
    \begin{tabular}{cc}
	\includegraphics[width=0.5\textwidth]{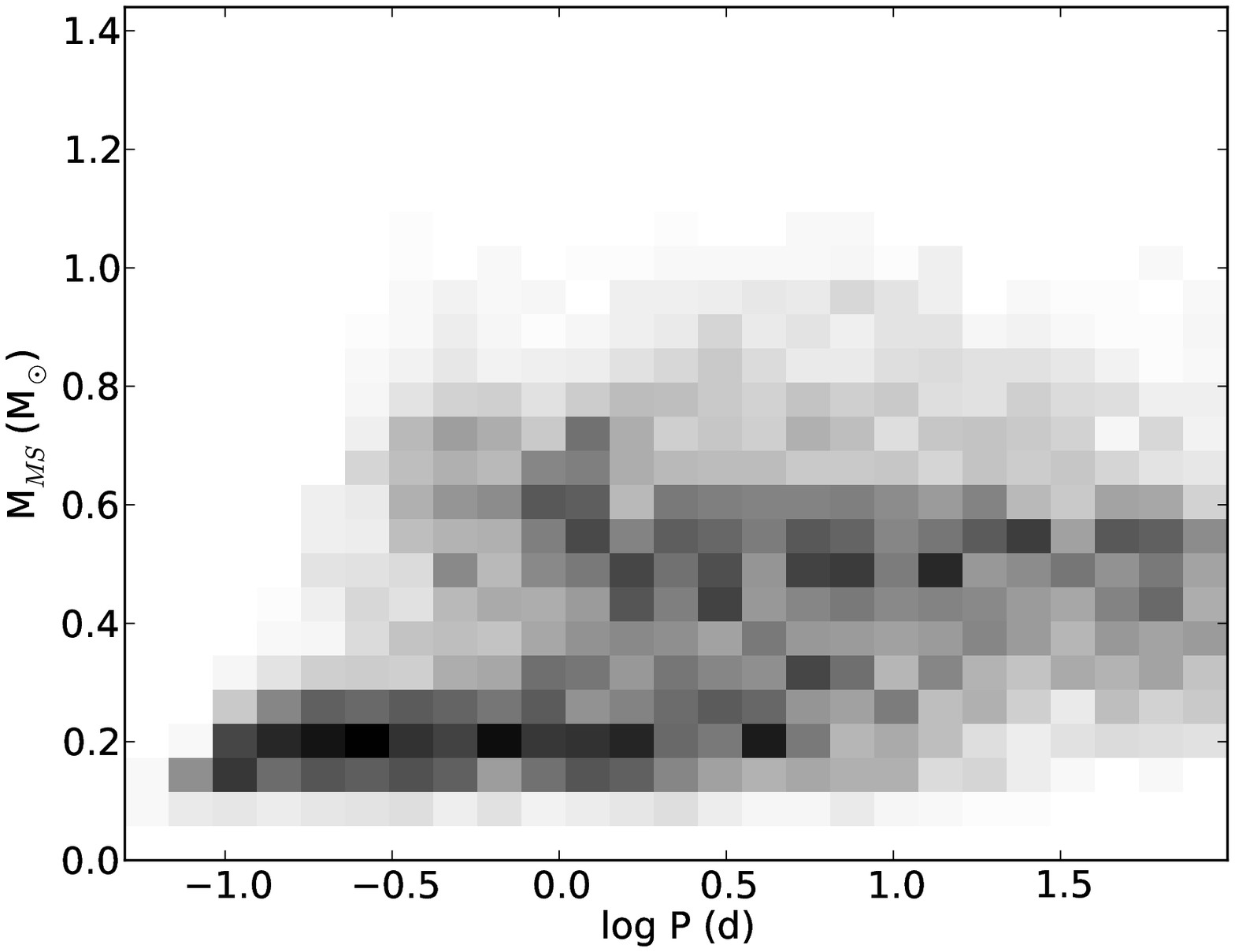} &
	\includegraphics[width=0.5\textwidth]{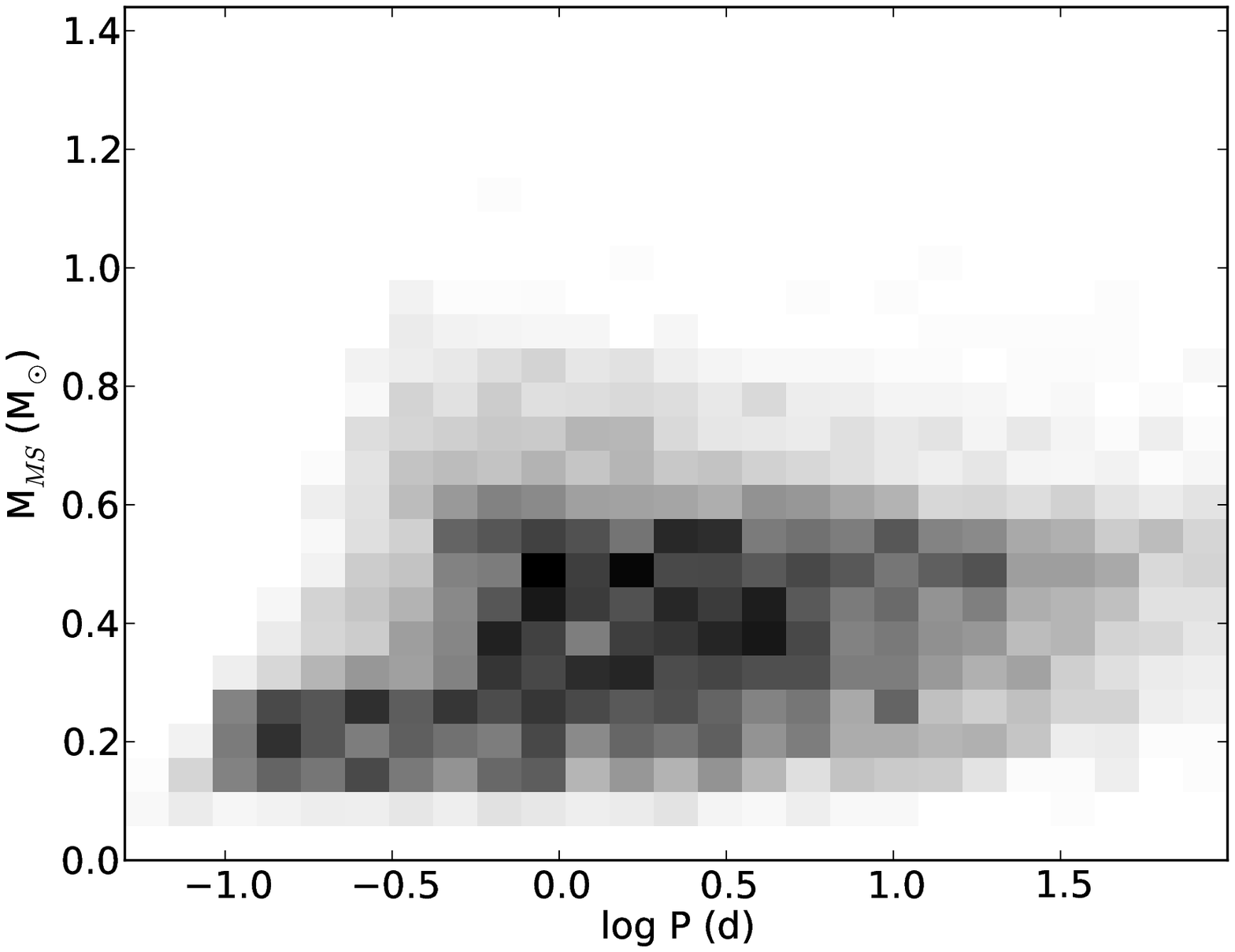} \\
	(a) & (b) \\
	\includegraphics[width=0.5\textwidth]{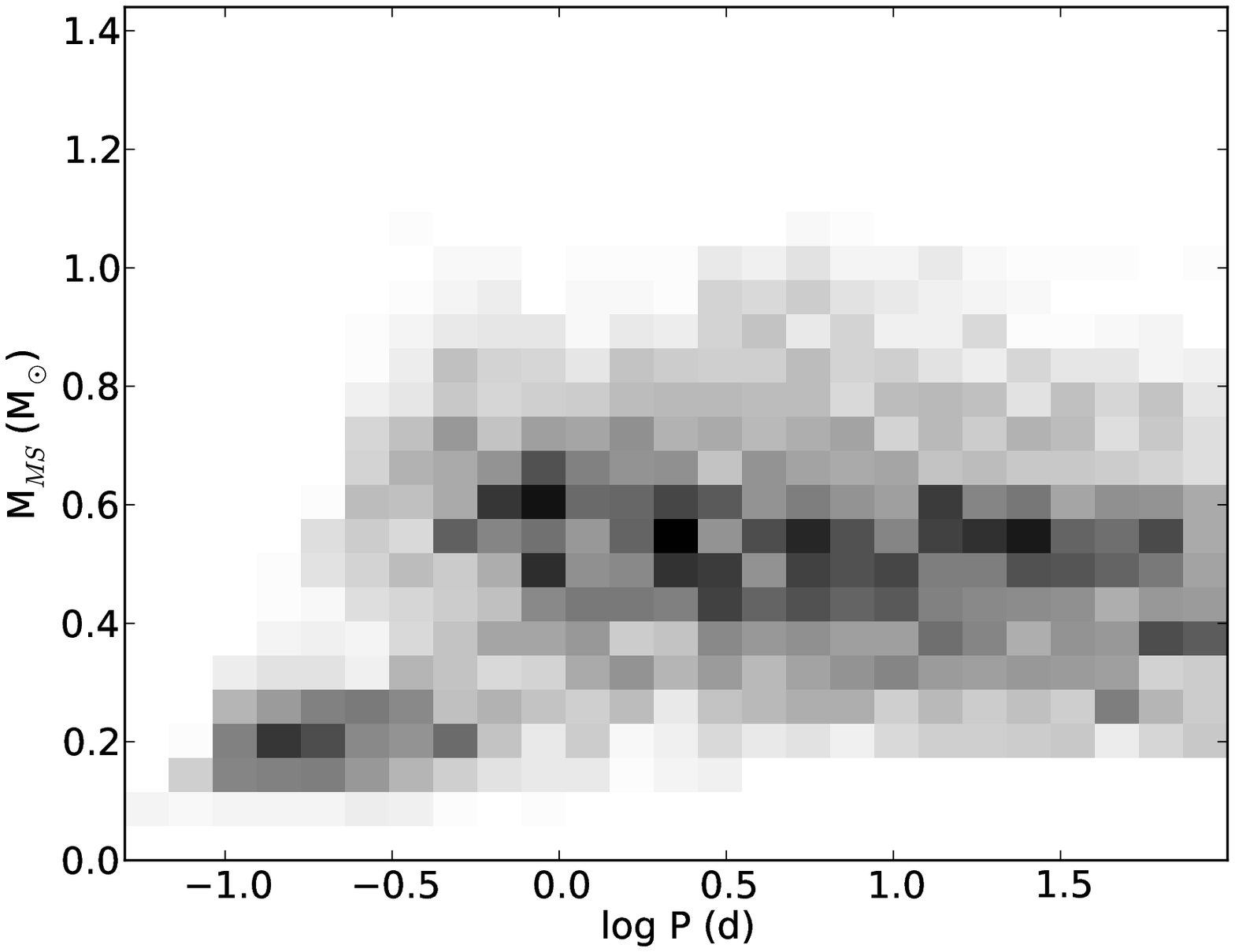} &
	\includegraphics[width=0.5\textwidth]{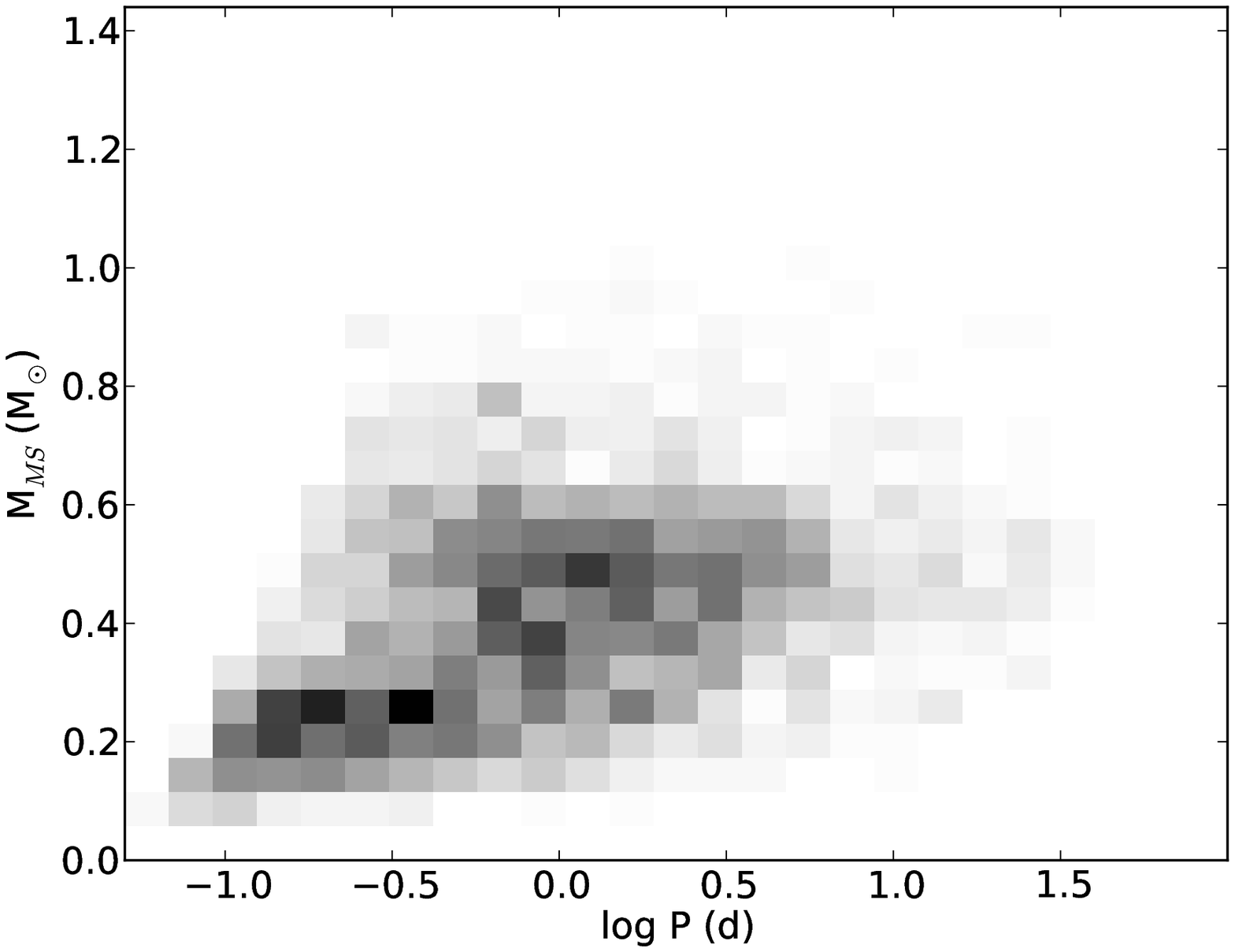} \\
	(c) & (d) \\
	\end{tabular}
    \caption{Visible population of PCEBs as a function of orbital period and mass of the MS star for all models: (a) model $\mga$1, (b) model $\maa$1, (c) model $\mga$2, and (d) model $\maa$2. The intensity of the grey scale corresponds to the density of objects on a linear scale.}
    \label{fig_ch4:pop_Mms_P}
    \end{figure*}

\begin{figure*}
    \centering
    \begin{tabular}{cc}
	\includegraphics[width=0.5\textwidth]{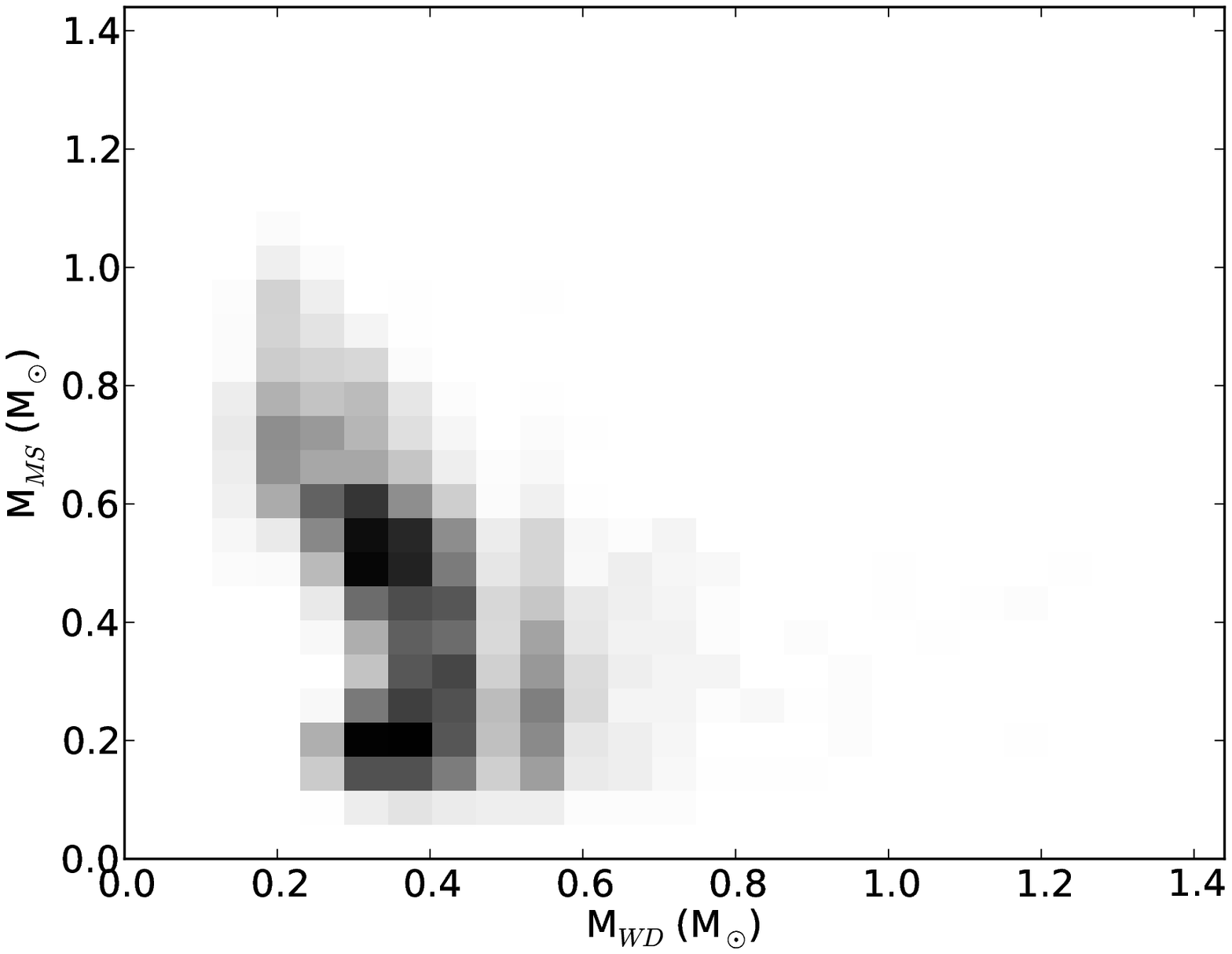} &
	\includegraphics[width=0.5\textwidth]{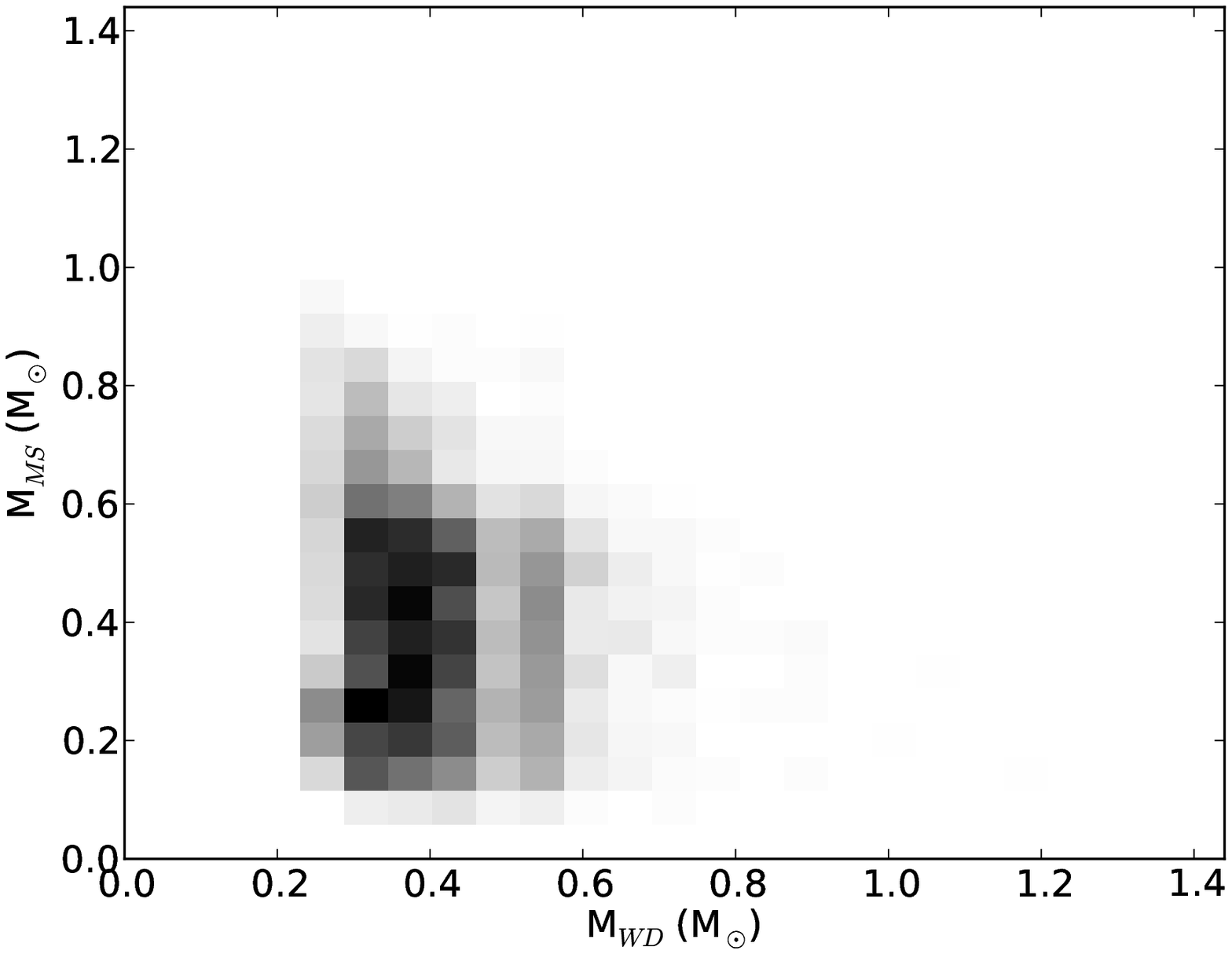} \\
	(a) & (b) \\
	\includegraphics[width=0.5\textwidth]{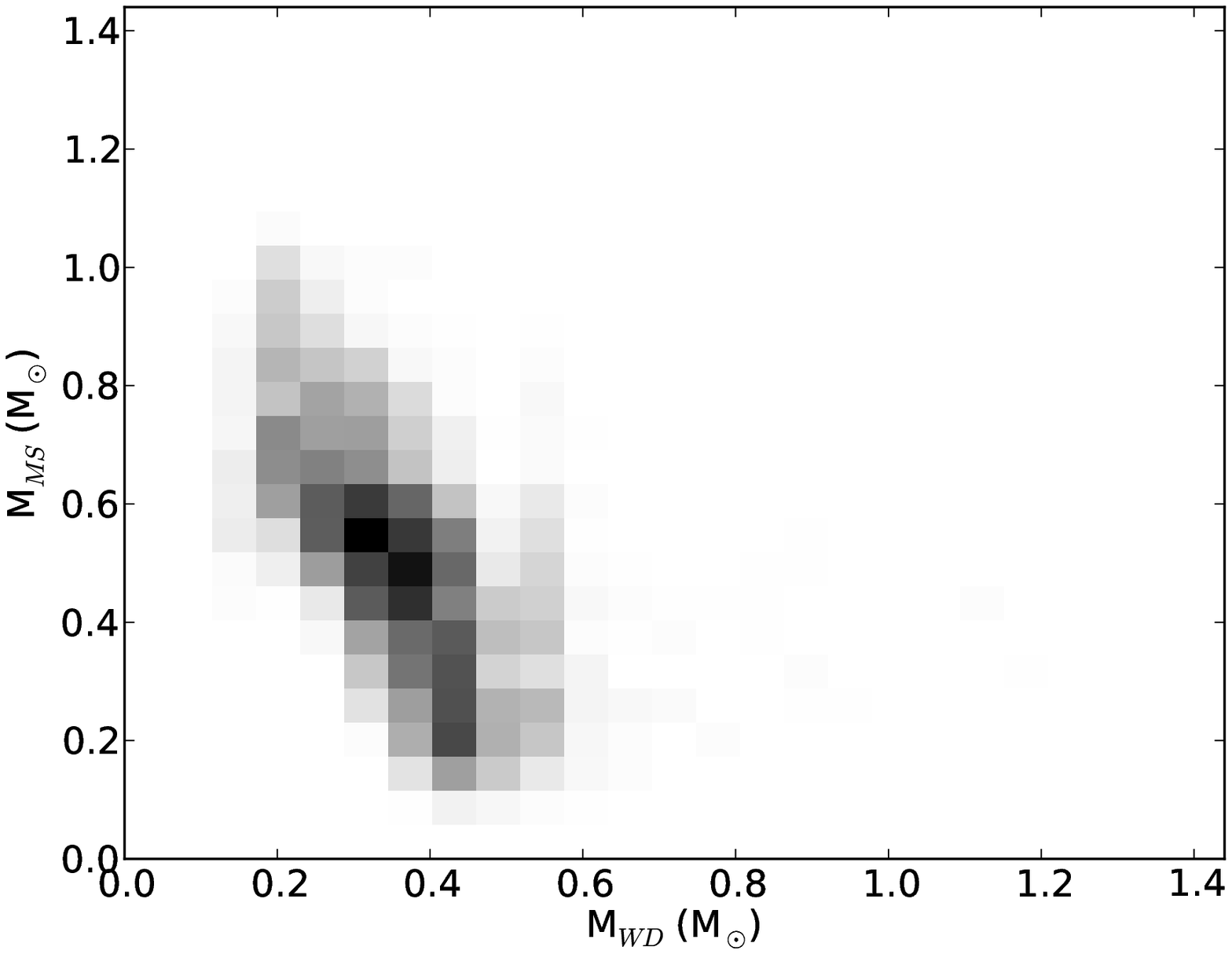} &
	\includegraphics[width=0.5\textwidth]{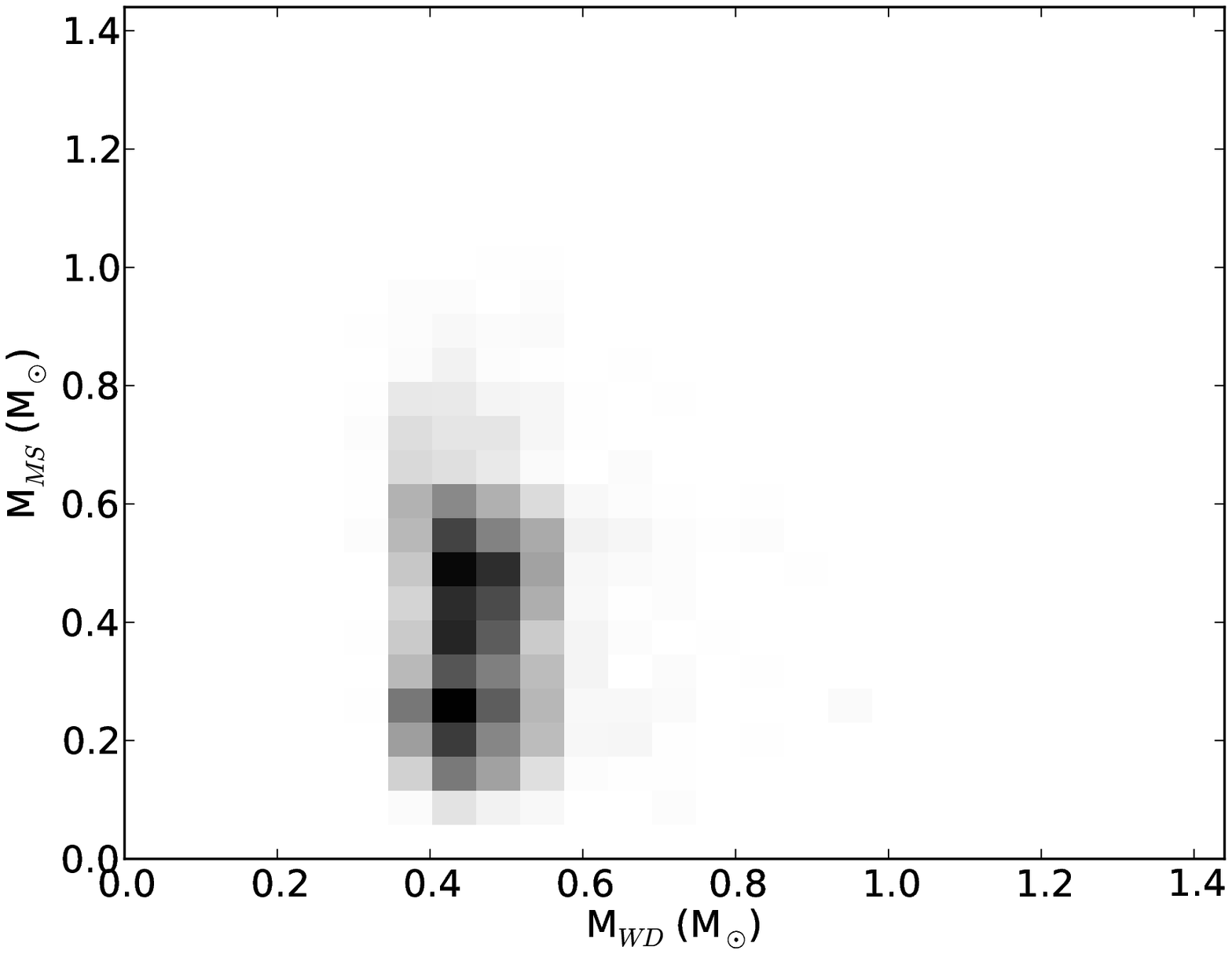} \\
	(c) & (d) \\
	\end{tabular}
    \caption{Visible population of PCEBs as a function of mass of the WD and the MS star for all models: (a) model $\mga$1, (b) model $\maa$1, (c) model $\mga$2, and (d) model $\maa$2. The intensity of the grey scale corresponds to the density of objects on a linear scale.}
    \label{fig_ch4:pop_Mms_Mwd}
    \end{figure*}

\begin{figure*}
    \centering
    \begin{tabular}{cc}
	\includegraphics[width=0.5\textwidth]{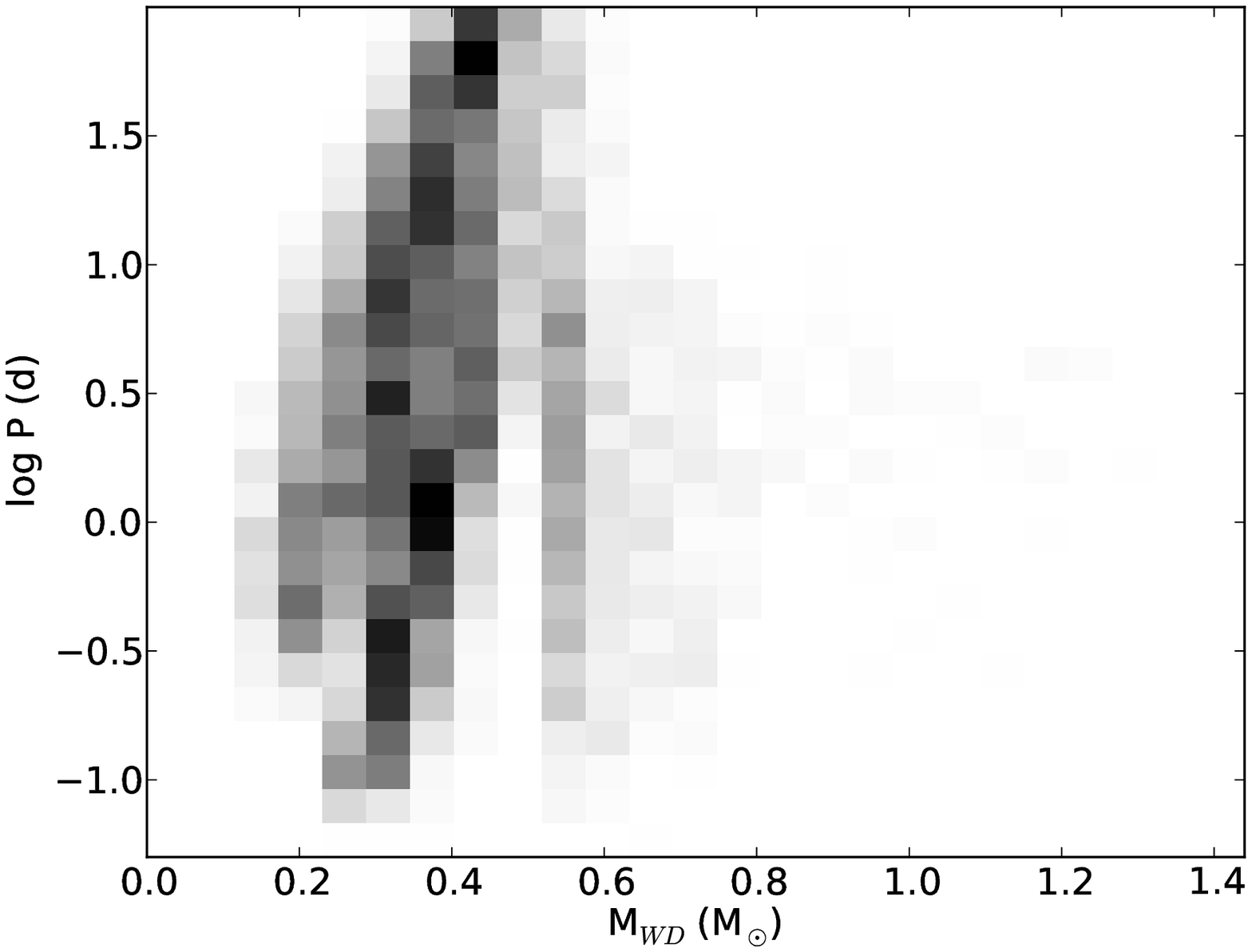} &
	\includegraphics[width=0.5\textwidth]{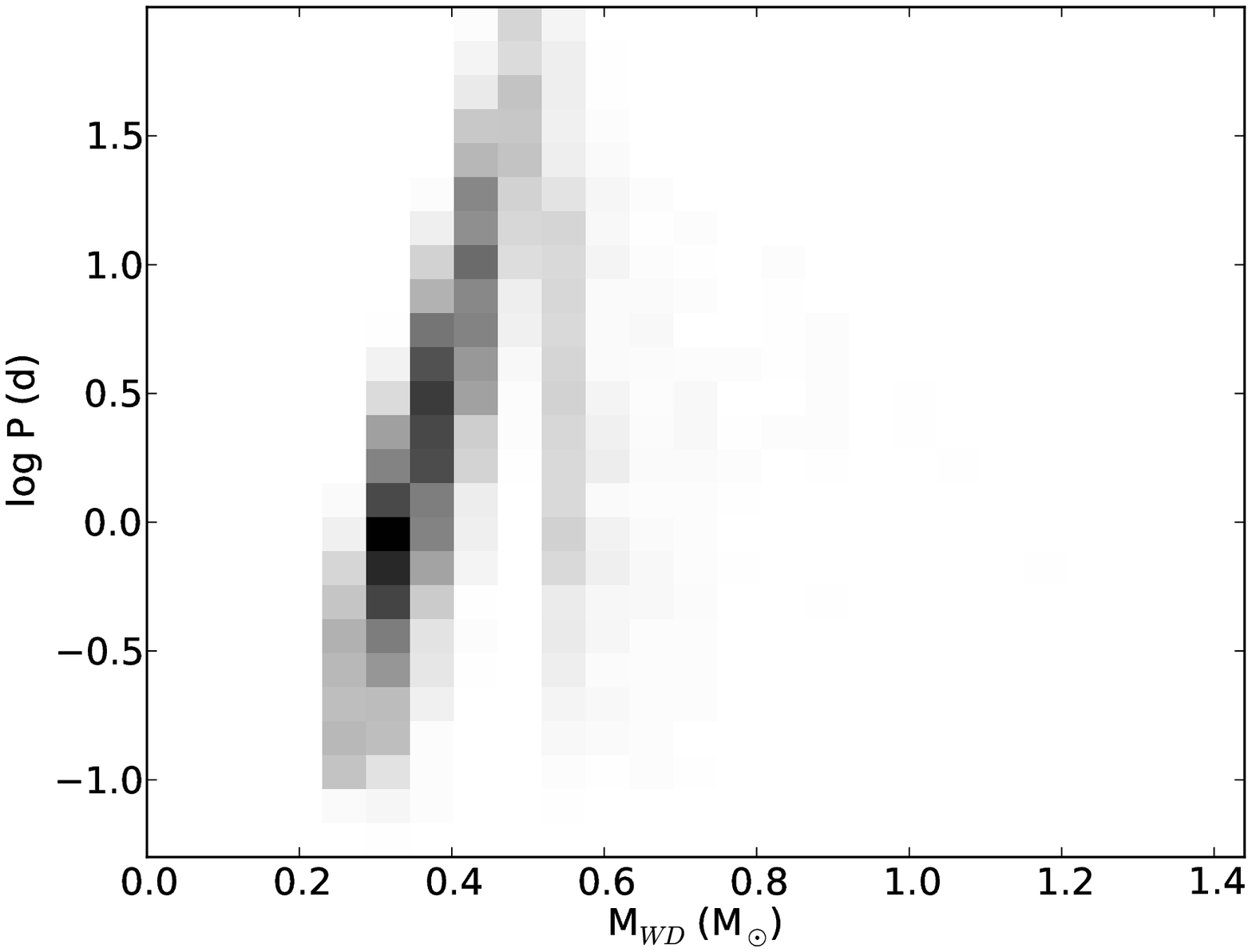} \\
	(a) & (b) \\
	\includegraphics[width=0.5\textwidth]{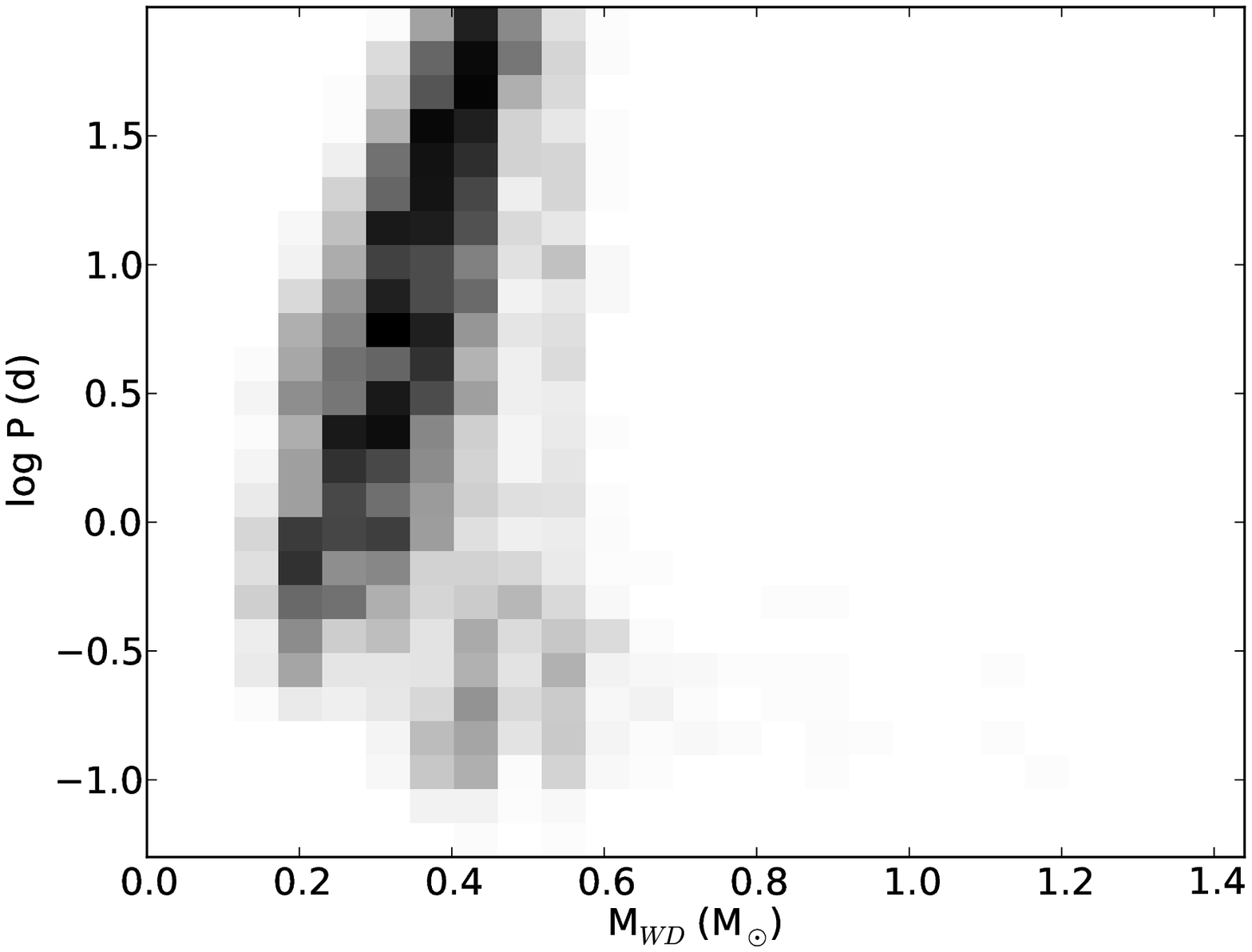} &
	\includegraphics[width=0.5\textwidth]{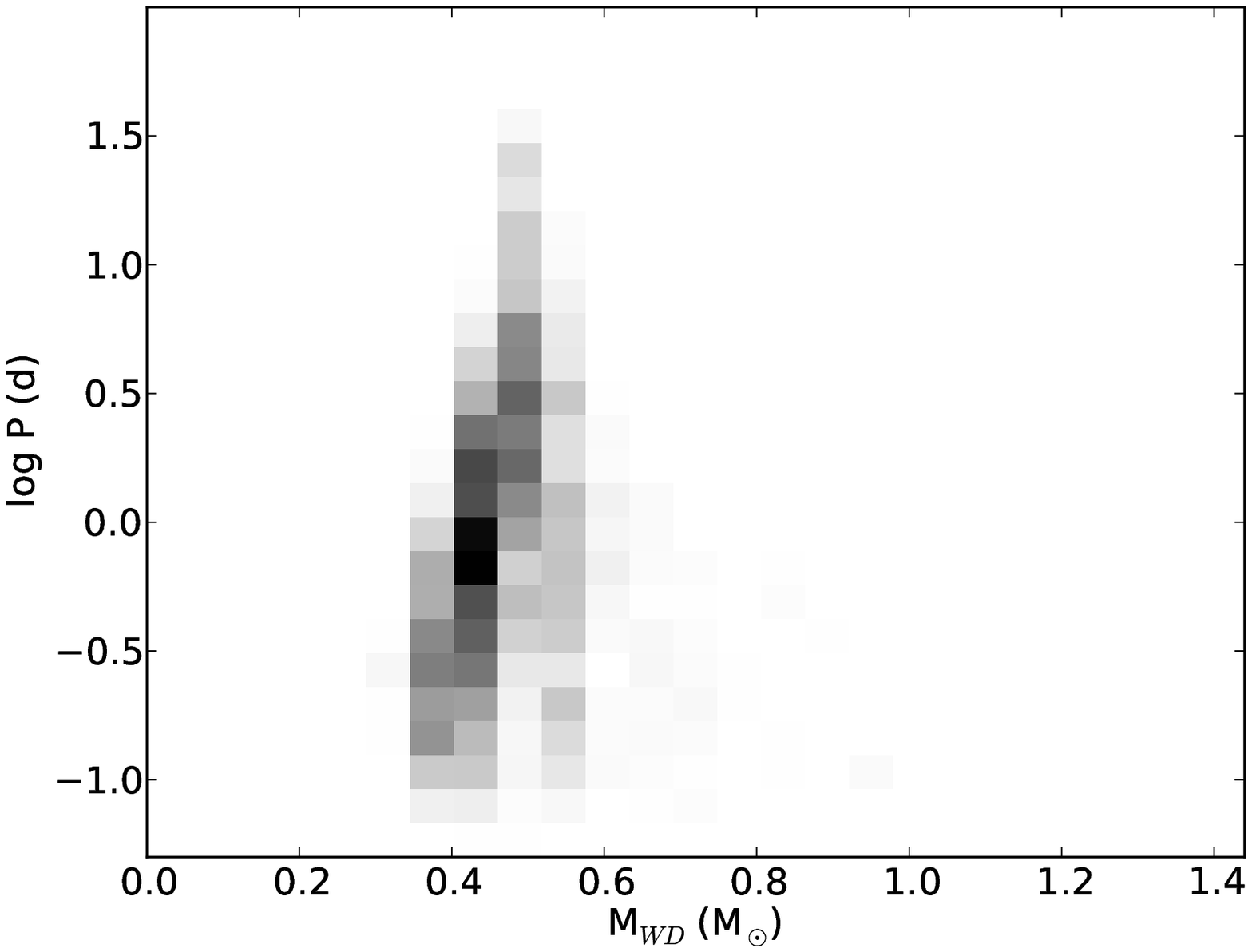} \\
	(c) & (d) \\
	\end{tabular}
    \caption{Visible population of PCEBs as a function of orbital period and WD mass for 
all models: (a) model $\mga$1, (b) model $\maa$1, (c) model $\mga$2, and (d) model $\maa$2. The intensity of the grey scale corresponds to the density of objects on a linear scale.}
    \label{fig_ch4:pop_P_Mwd}
    \end{figure*}

Figures\,\ref{fig_ch4:pop_Mms_P},~\ref{fig_ch4:pop_Mms_Mwd},~and~\ref{fig_ch4:pop_P_Mwd} show the distribution of MS mass, WD mass, and orbital period of the visible population of PCEBs. Model $\mga$1, $\maa$1 and $\mga$2 show PCEB systems with periods between 0.05-100d, whereas model $\maa$2 shows a narrower period range of about 0.05-10d. 
Few PCEBs exist at periods of less than a few hours, as these systems come in contact and possibly evolve into CVs. 
Figure\,\ref{fig_ch4:pop_Mms_Mwd} shows a relation between MS and WD mass that is different for each model. 
The masses of WDs in visible PCEBs are roughly between 0.2 and 0.8\Msolar; most WDs have either helium (He) or carbon-oxygen (CO) cores.
Figure\,\ref{fig_ch4:pop_P_Mwd} shows that the model $\mga$1 and model $\mga$2 periods at a given WD mass can be longer than for model $\maa$1 and model $\maa2$. This is because the CE phase leads to a strong decrease in the orbital separation according to the $\alpha$-prescription, while this is not necessarily true in the $\gamma$-prescription. 

Varying the cuts that determine which PCEBs are visible (see Sect.\,\ref{sec_ch4:vis}), does not change our results much. The limiting magnitude of $g=15-20$ does not affect the relations between WD mass, MS mass, and period, but it can effect the space density of visible PCEBs. If the sensitivity of the observations increases to $g=21$, the space density within 200pc and 500pc increases by about 15-30\% and 30-50\% respectively. 

The cut that distinguishes WDMS from apparent single WDs (see eq.\,\ref{eq_ch4:dz}) has a small effect on the population of PCEBs. If we assume a more conservative cut, less massive MS stars are visible at a given WD temperature. Varying the cut between $\Delta z > 0$ and $\Delta z > -2$ does not affect the space density significantly. The distribution of periods is not affected.
Making a cut in the i-band instead of the z-band has a similar effect on the visible PCEB population as varying $\Delta z$.  

Varying the cut that distinguishes WDMS from apparent single MS stars (see eq.\,\ref{eq_ch4:dg}) is important for the space density and the MS mass distribution of PCEBs, as the WD is less bright than the secondary star for most systems. If a more conservative cut is appropriate, i.e. $\Delta g > 0$, the space density decreases by about 30-40\%, and the less massive MS stars are visible at a given WD temperature. 
The space density increases by 40-50\% when assuming $\Delta g > 2$.  
Varying the cut between $\Delta g > 0$ and $\Delta g > 2$ affects the maximum MS mass in PCEBs by $\pm$0.1\Msolar. 
Most importantly, the correlations of MS mass with WD mass and period are, however, not affected. The effect on the visible PCEB population of making a cut in the u-band is comparable to that in the g-band.

\subsection{The SDSS PCEB sample}
\label{sec_ch4:res_sdss} 
\begin{figure*}
\centering
\begin{tabular}{cccc}
\includegraphics[width=0.3\textwidth]{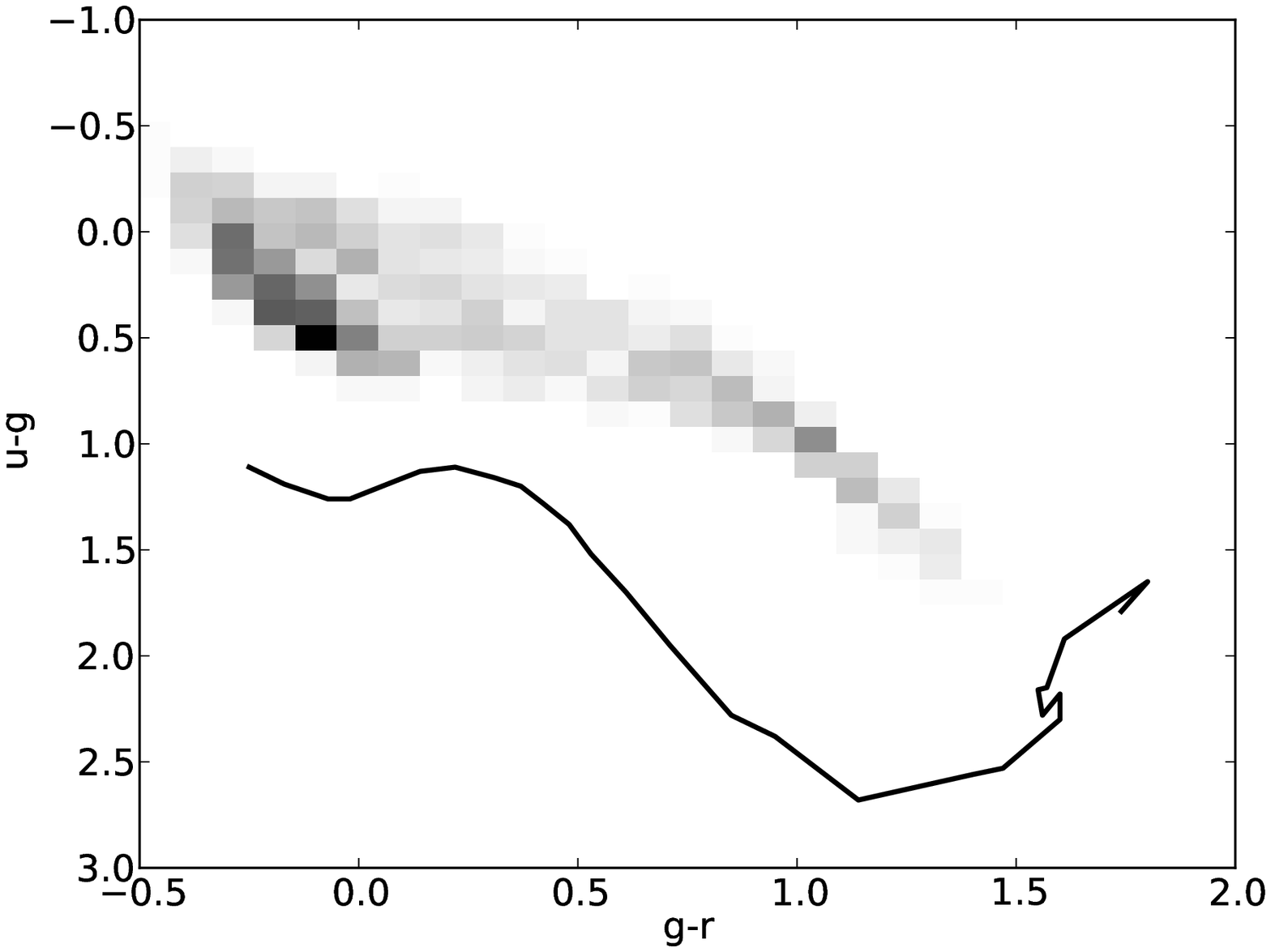} &
\includegraphics[width=0.3\textwidth]{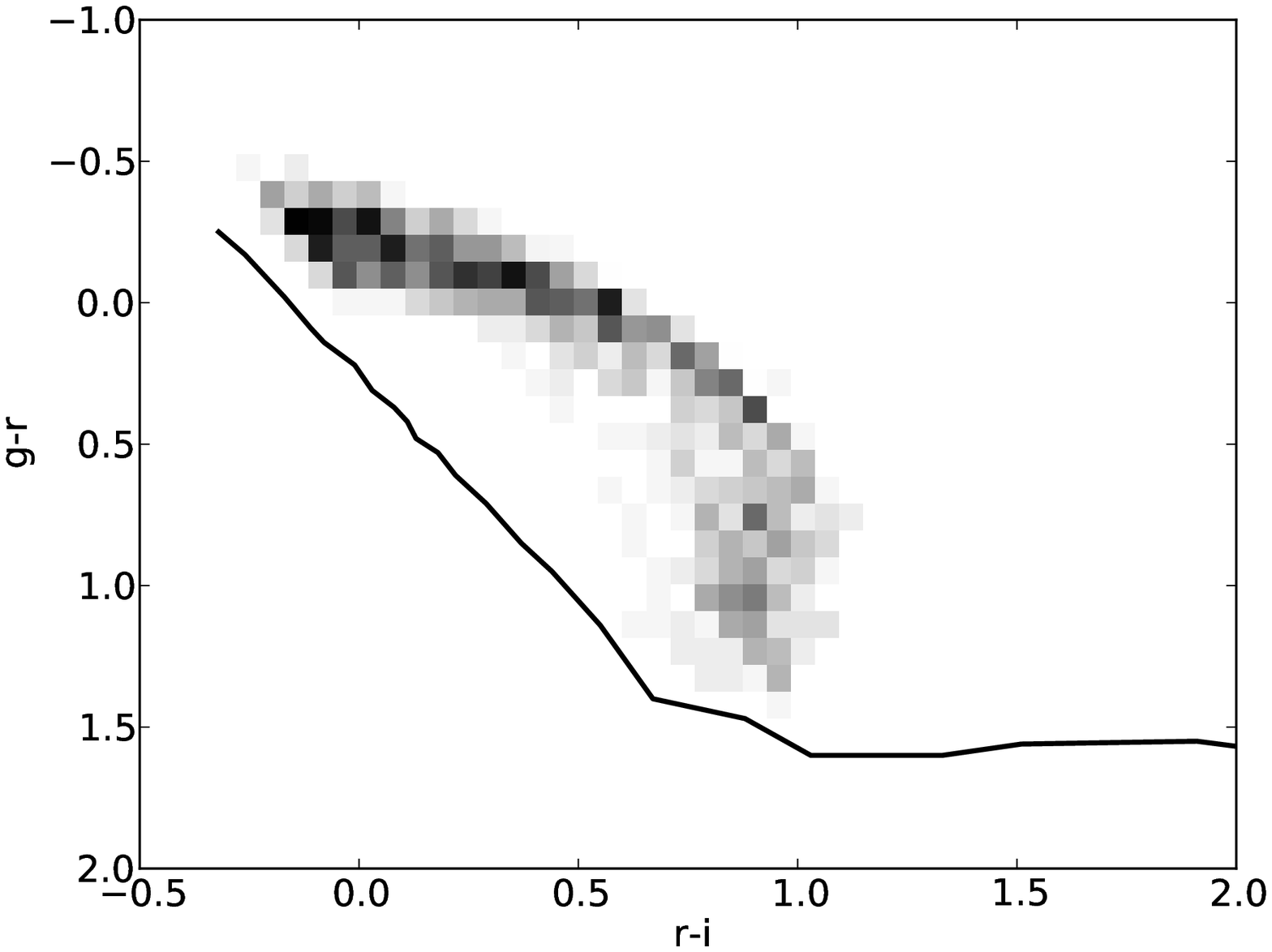} & 
\includegraphics[width=0.3\textwidth]{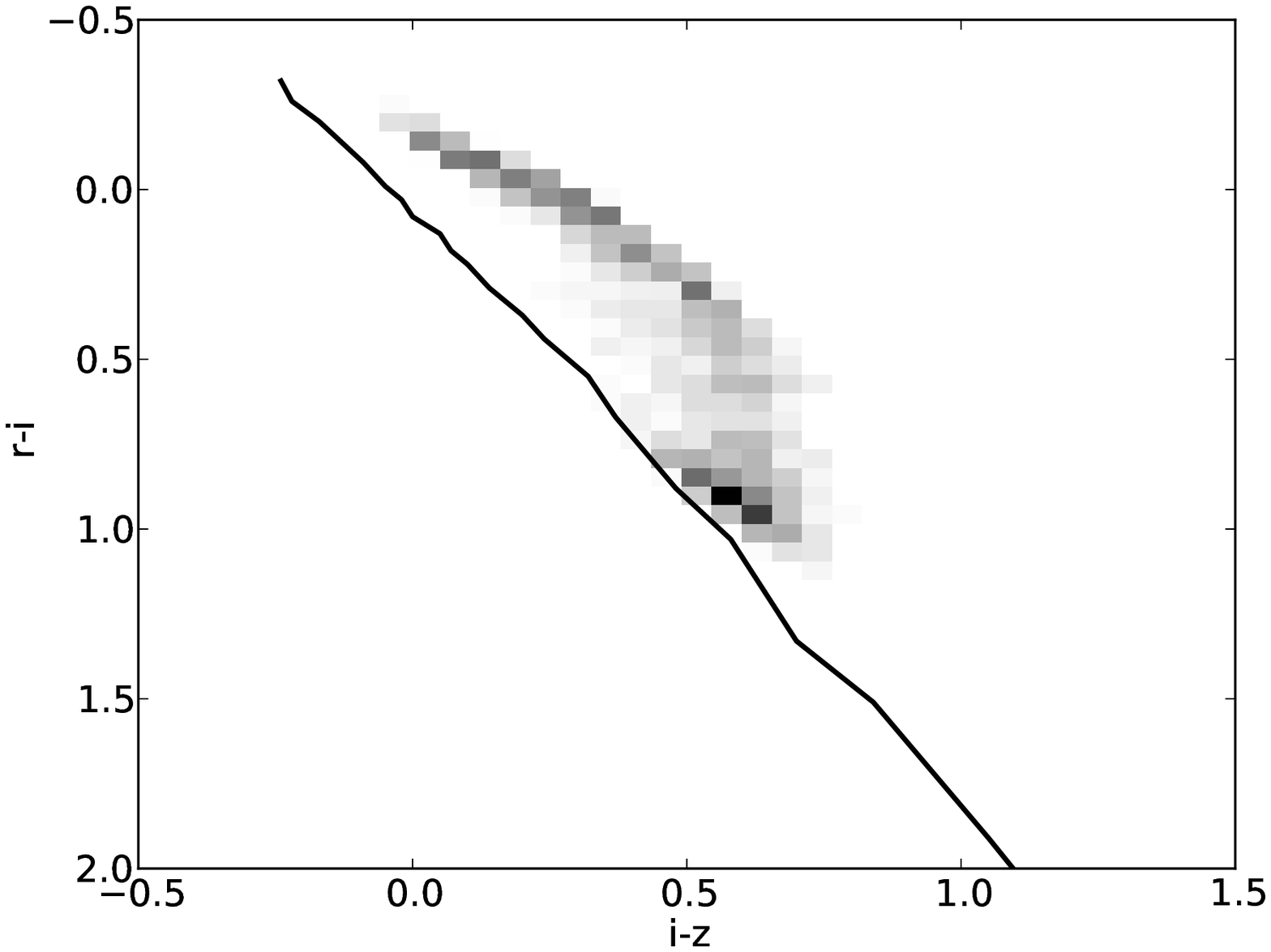} \\
\end{tabular}
    \caption{Color-color diagrams for the visible population of PCEBs in the SDSS for model $\maa$2. On the left, it shows the u-g vs. g-r diagram, in the middle, the g-r vs. r-i diagram, and on the right, the r-i vs. i-z diagram. The intensity of the grey scale corresponds to the density of objects on a linear scale. The solid line corresponds to the unreddened MS from A-type to M-type MS stars. The color-color diagrams are very comparable to those of model $\maa$1, model $\mga$1, and model $\mga$2. } 
\label{fig_ch4:color_color_sdss}
\end{figure*}

   \begin{figure*}
    \centering
    \begin{tabular}{cc}
	\includegraphics[width=0.5\textwidth]{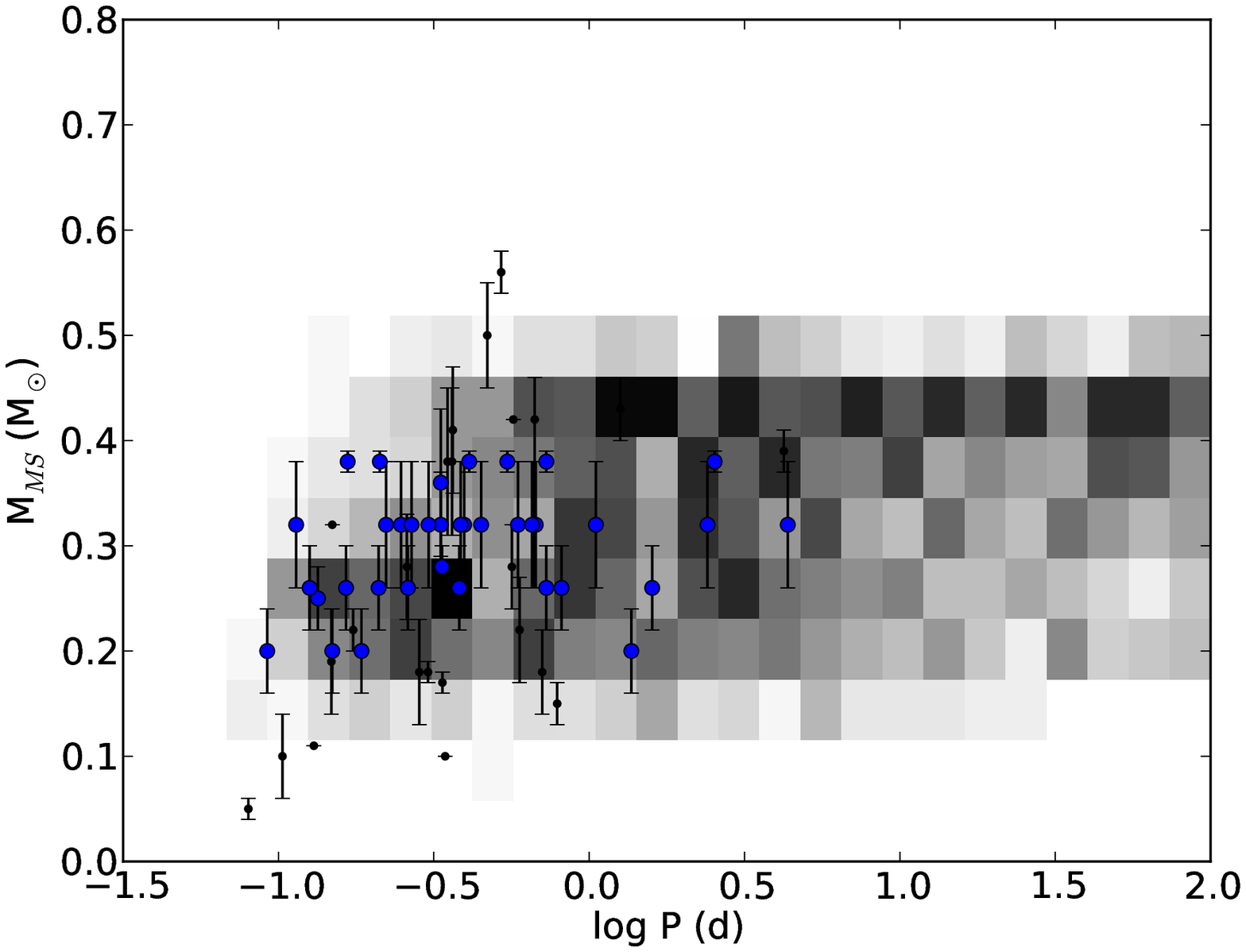} &
	\includegraphics[width=0.5\textwidth]{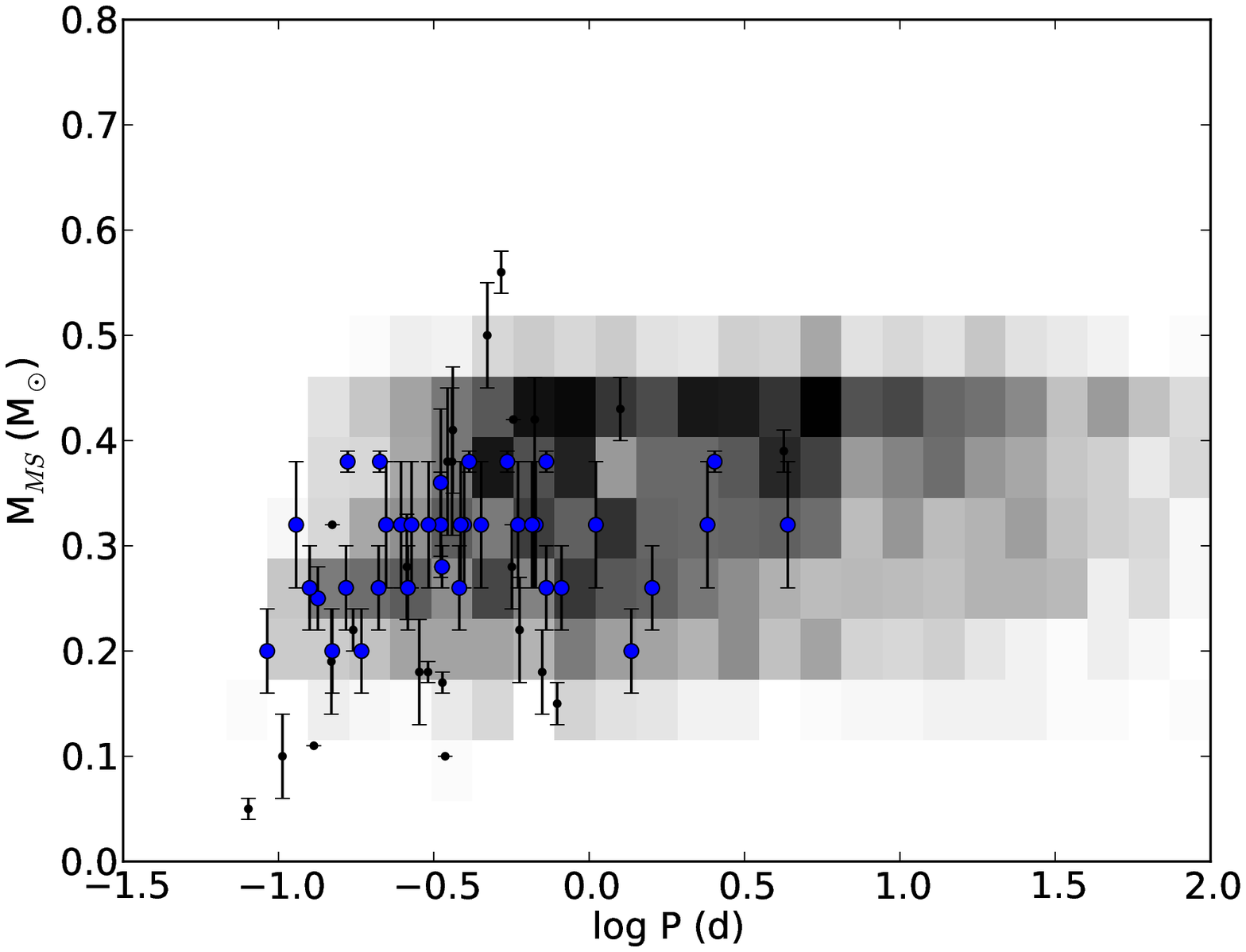} \\
	(a) & (b) \\
	\includegraphics[width=0.5\textwidth]{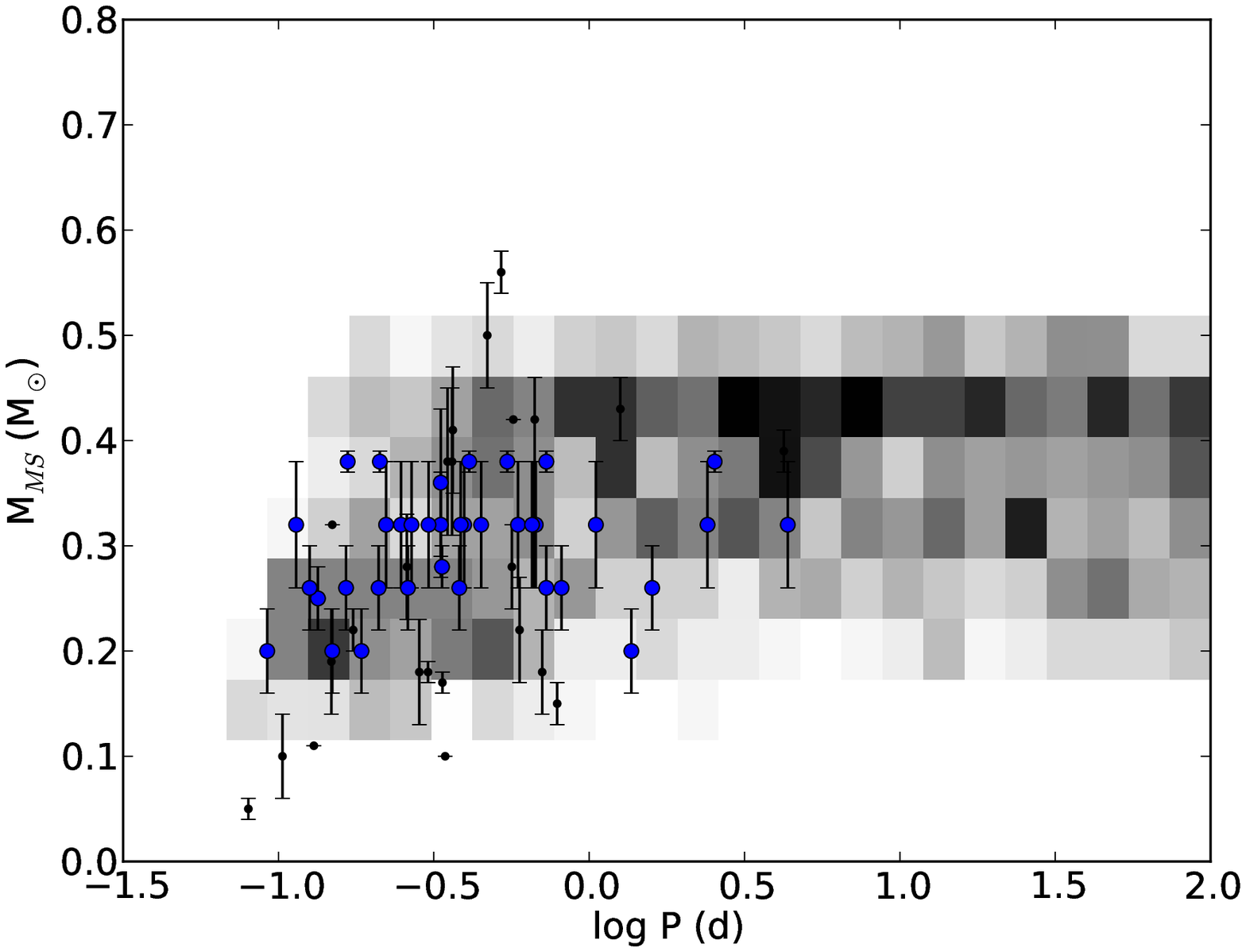} &
	\includegraphics[width=0.5\textwidth]{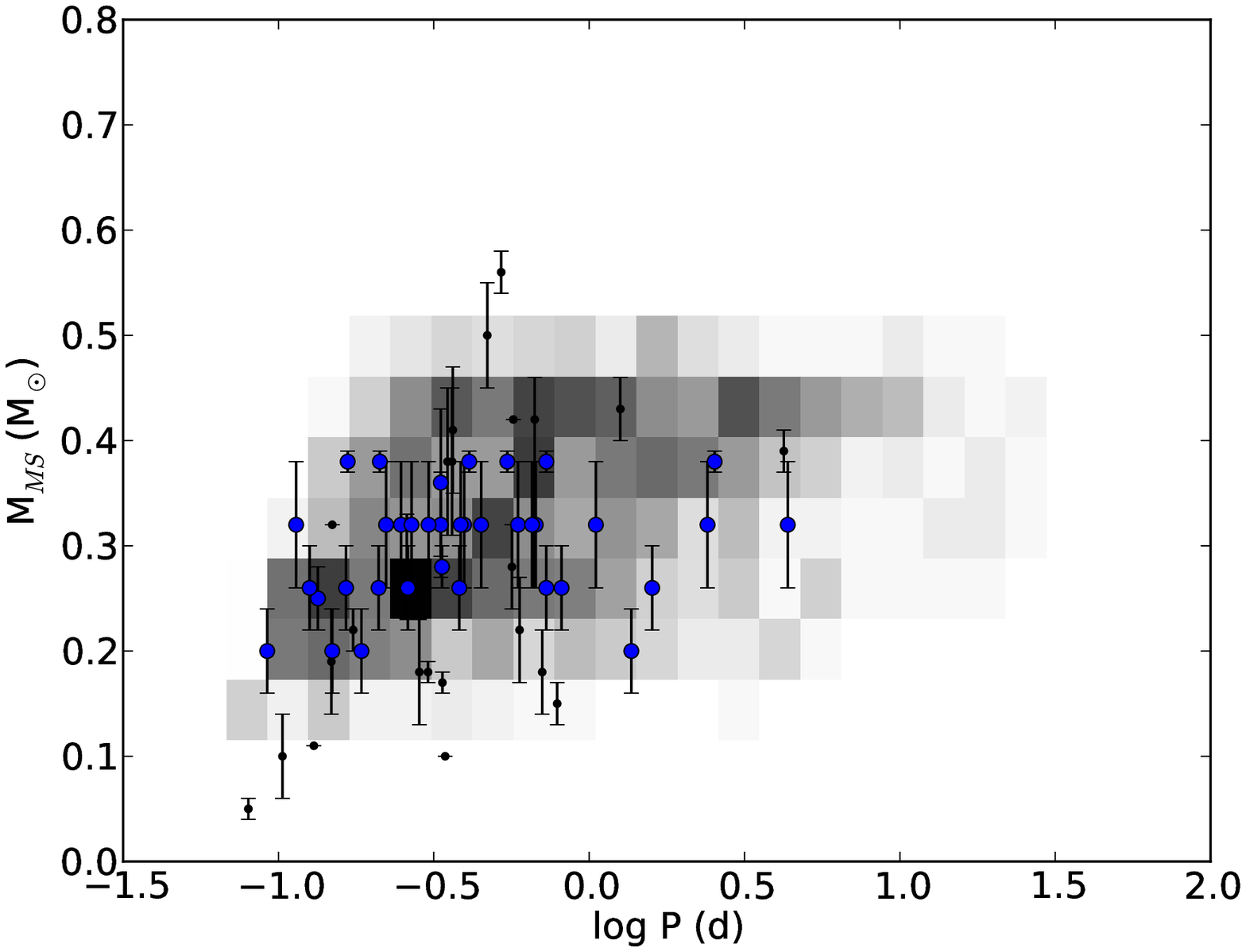} \\
	(c) & (d) \\
	\end{tabular}
    \caption{Visible population of PCEBs in the SDSS as a function of orbital period and mass of the MS star for all models: (a) model $\mga$1, (b) model $\maa$1, (c) model $\mga$2, and (d) model $\maa$2. The intensity of the grey scale corresponds to the density of objects on a linear scale. Overplotted are the observed PCEBs taken from \citet{Zor11b}. Thick points represent systems that are found by the SDSS, and thin points represent previously known PCEBs with accurately measured parameters. The previously known sample of PCEBs is affected by other selection effects than the SDSS sample or the synthetic sample. Note that Ik Peg has been removed from the sample as its MS component is not an M-dwarf. }
    \label{fig_ch4:pop_Mms_P_sdss}
    \end{figure*}

    \begin{figure*}
    \centering
    \begin{tabular}{cc}
	\includegraphics[width=0.5\textwidth]{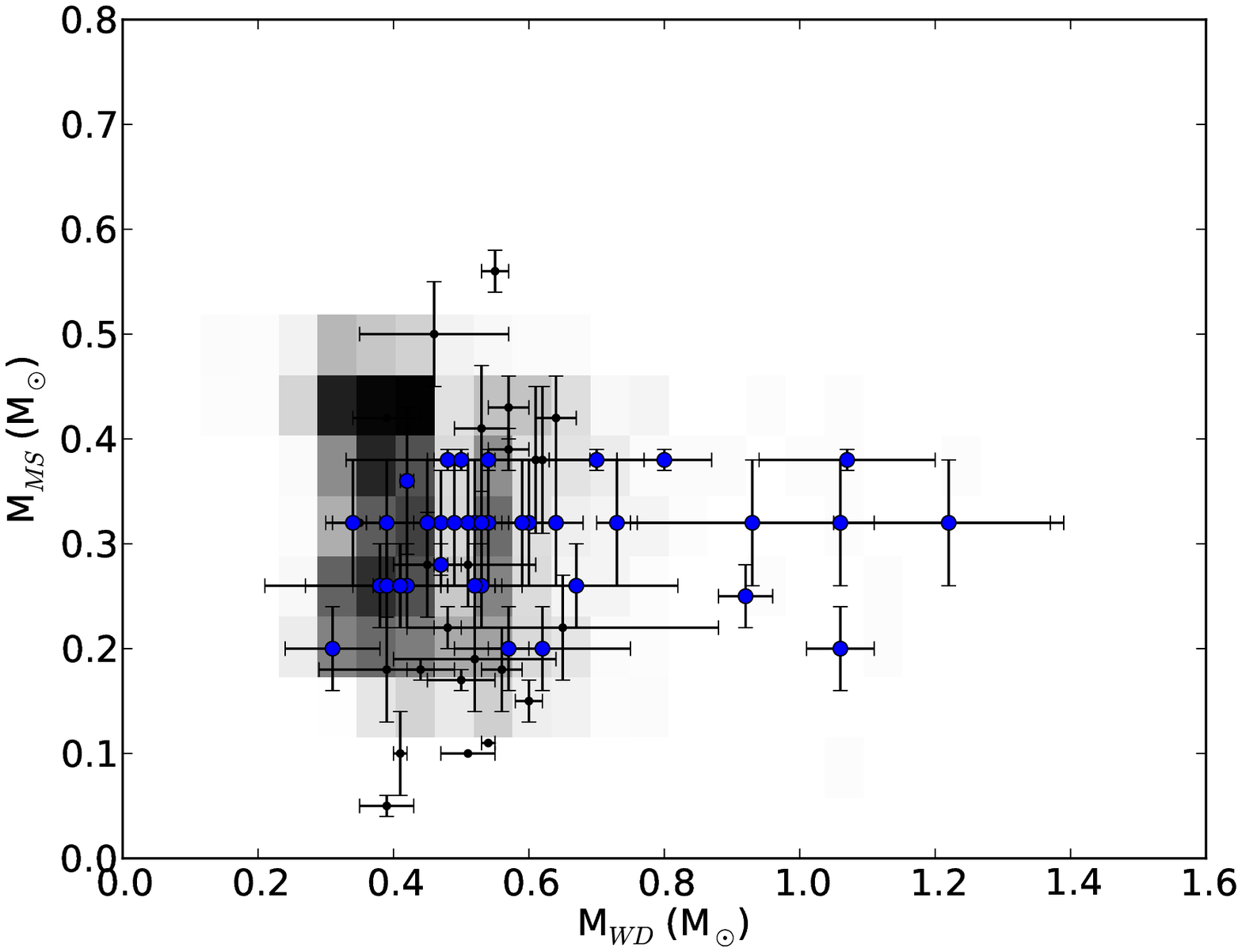} &
	\includegraphics[width=0.5\textwidth]{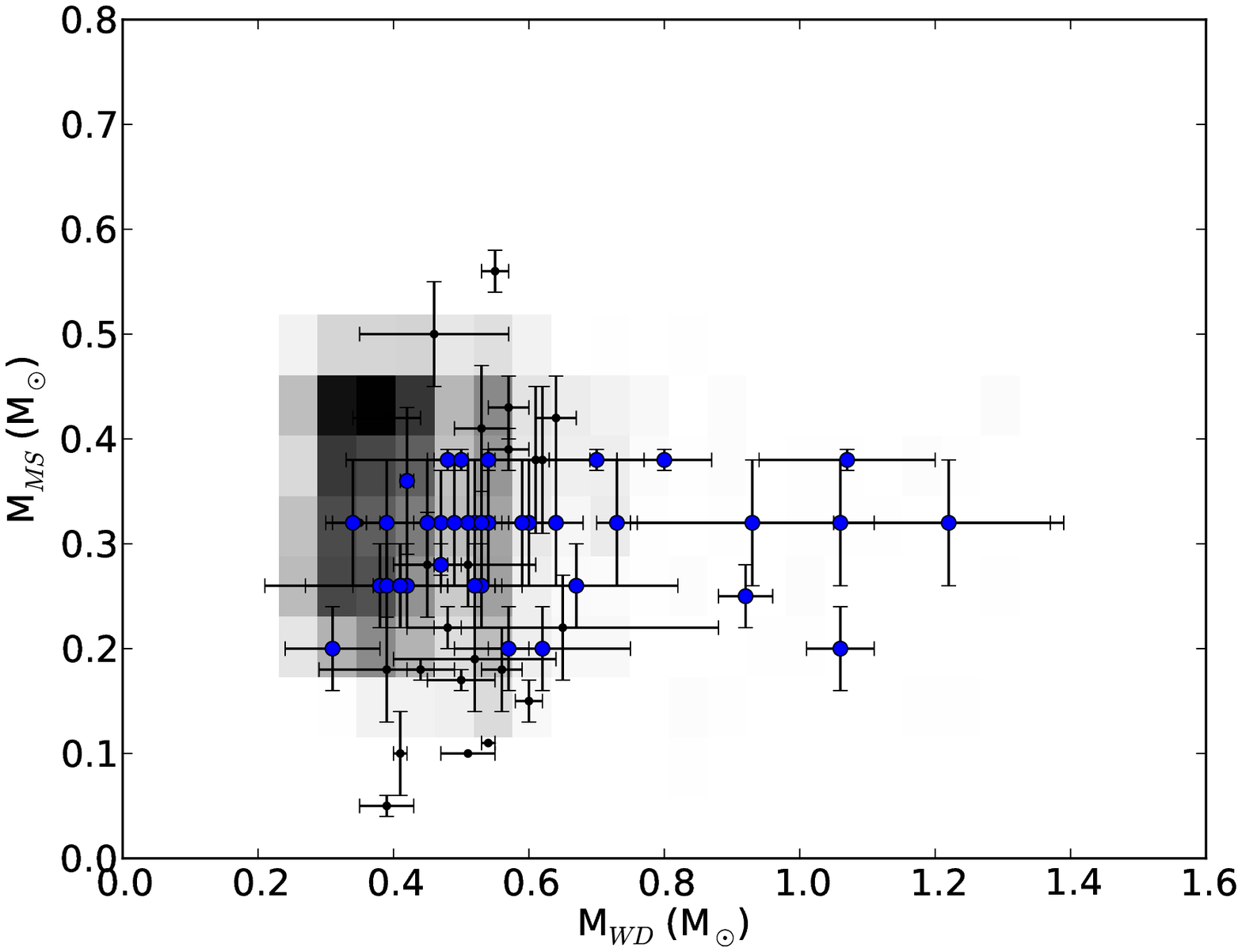} \\
	(a) & (b) \\
	\includegraphics[width=0.5\textwidth]{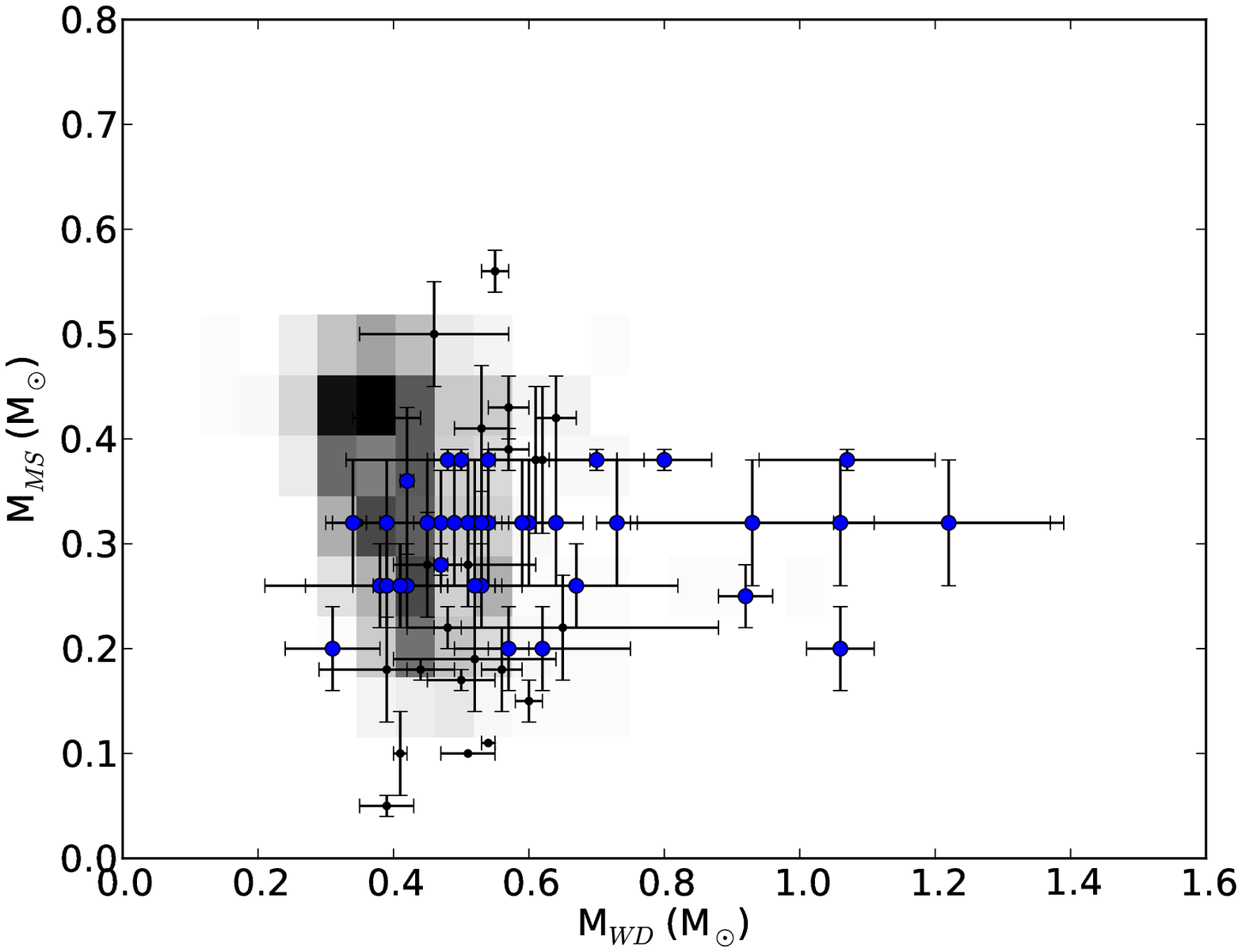} &
	\includegraphics[width=0.5\textwidth]{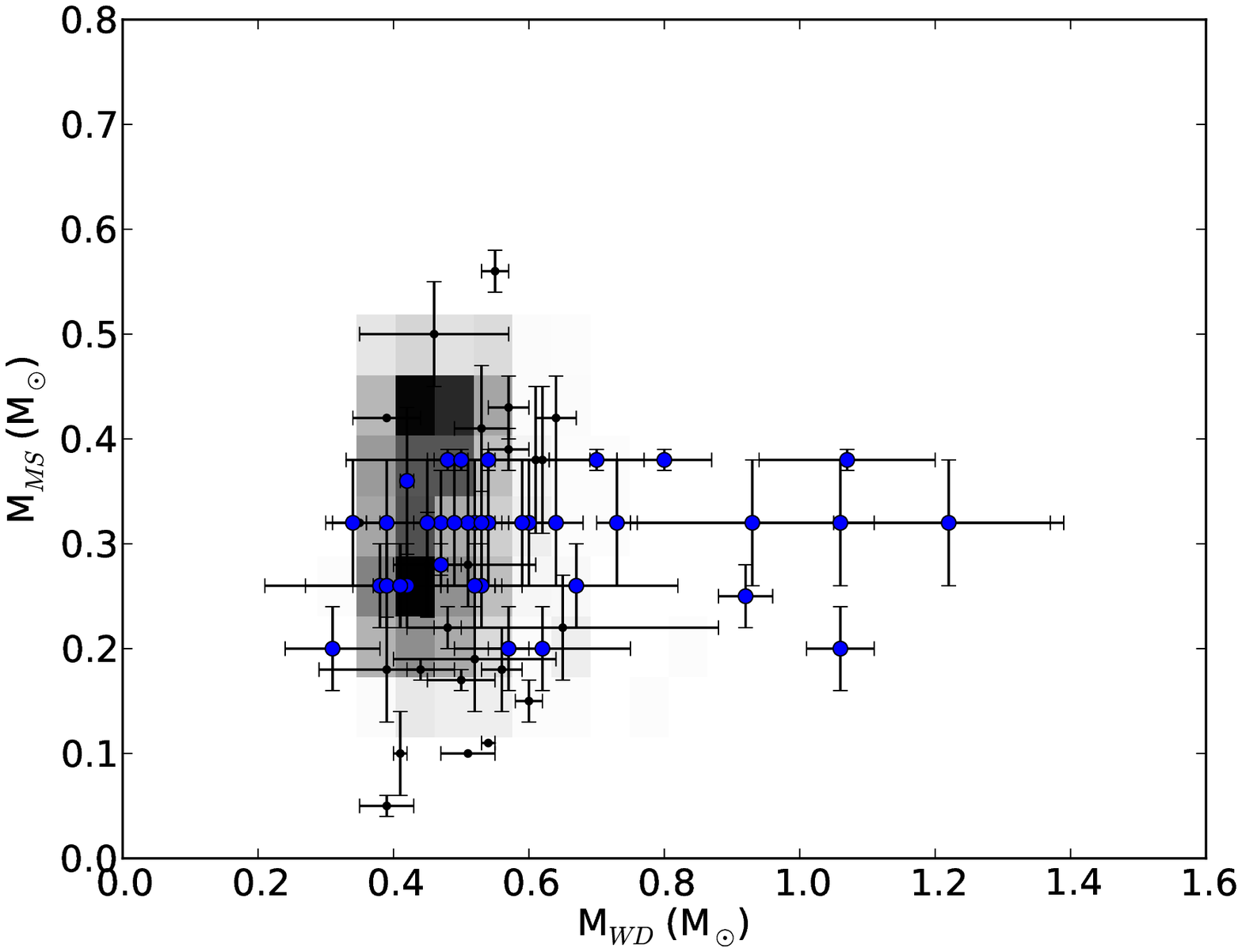} \\
	(c) & (d) \\
	\end{tabular}
    \caption{Visible population of PCEBs in the SDSS as a function of mass of the WD and the MS star for all models: (a) model $\mga$1, (b) model $\maa$1, (c) model $\mga$2, and (d) model $\maa$2. The intensity of the grey scale corresponds to the density of objects on a linear scale. Overplotted are the observed PCEBs taken from \citet{Zor11b}. Thick points represent systems that are found by the SDSS, and thin points represent previously known PCEBs with accurately measured parameters. The previously known sample of PCEBs is affected by other selection effects than the SDSS sample or the synthetic sample. Note that Ik Peg has been removed from the sample as its MS component  is not an M-dwarf. }
    \label{fig_ch4:pop_Mms_Mwd_sdss}
    \end{figure*}

    \begin{figure*}
    \centering
    \begin{tabular}{cc}
	\includegraphics[width=0.5\textwidth]{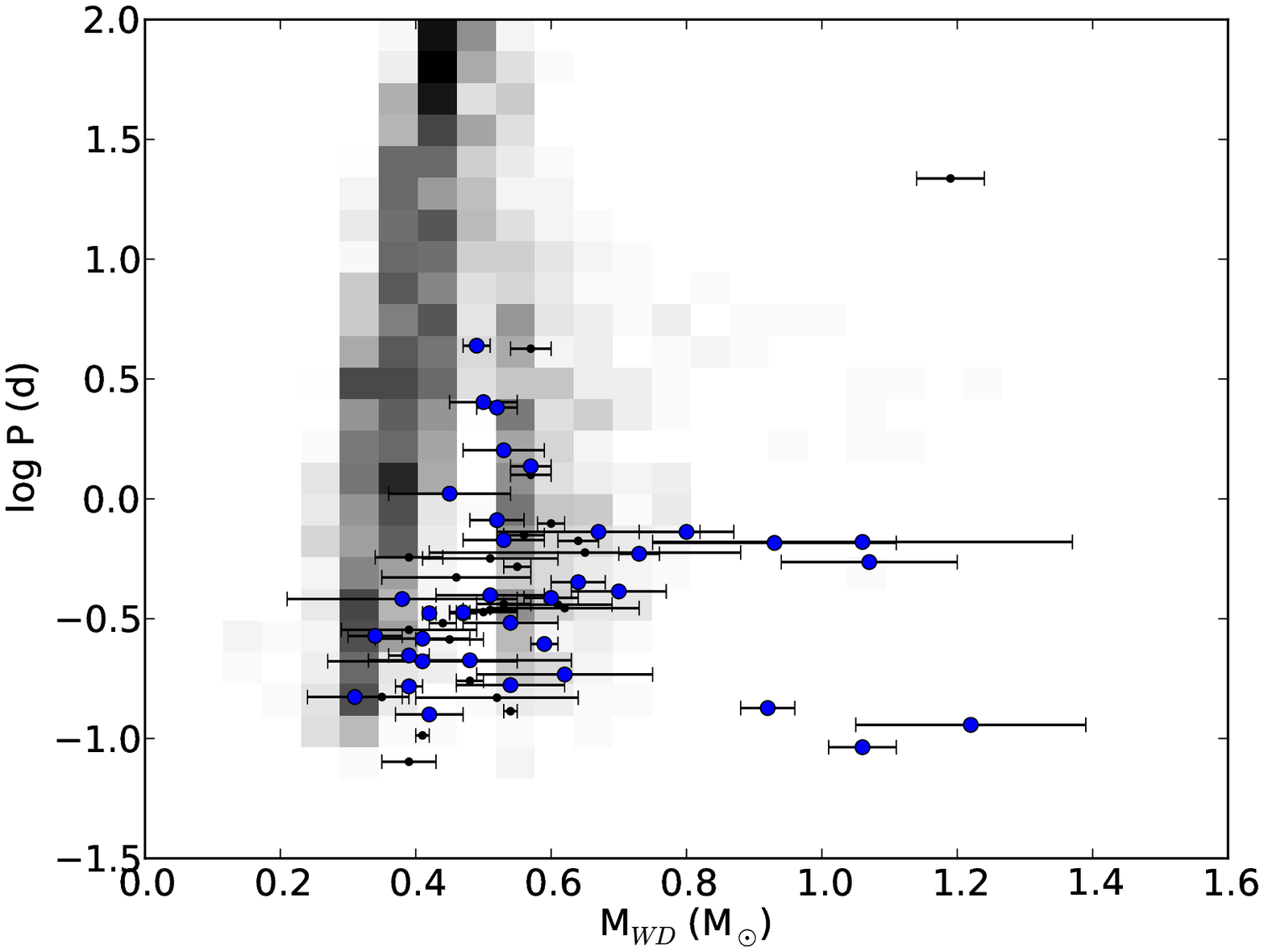} &
	\includegraphics[width=0.5\textwidth]{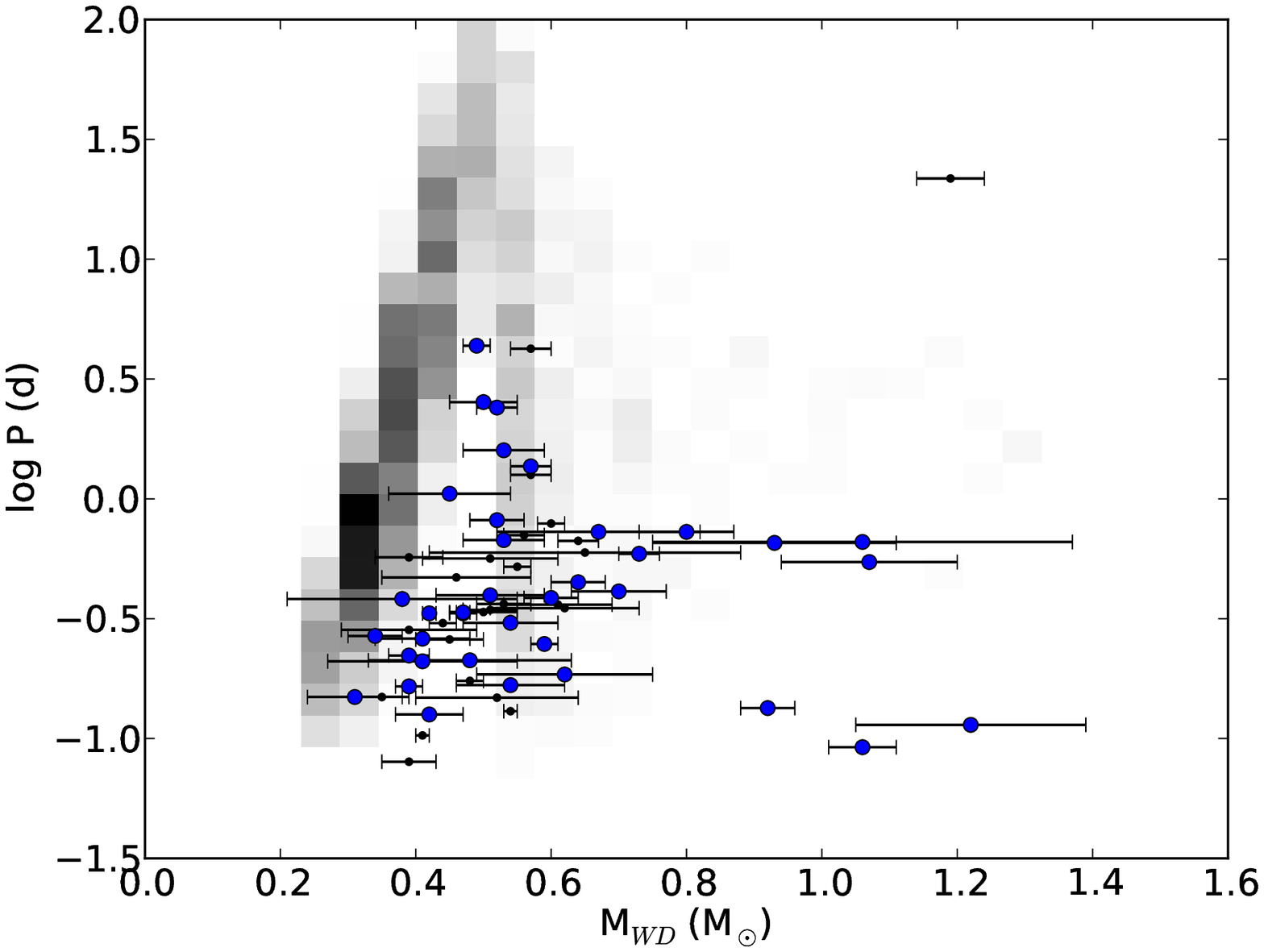} \\
	(a) & (b) \\
	\includegraphics[width=0.5\textwidth]{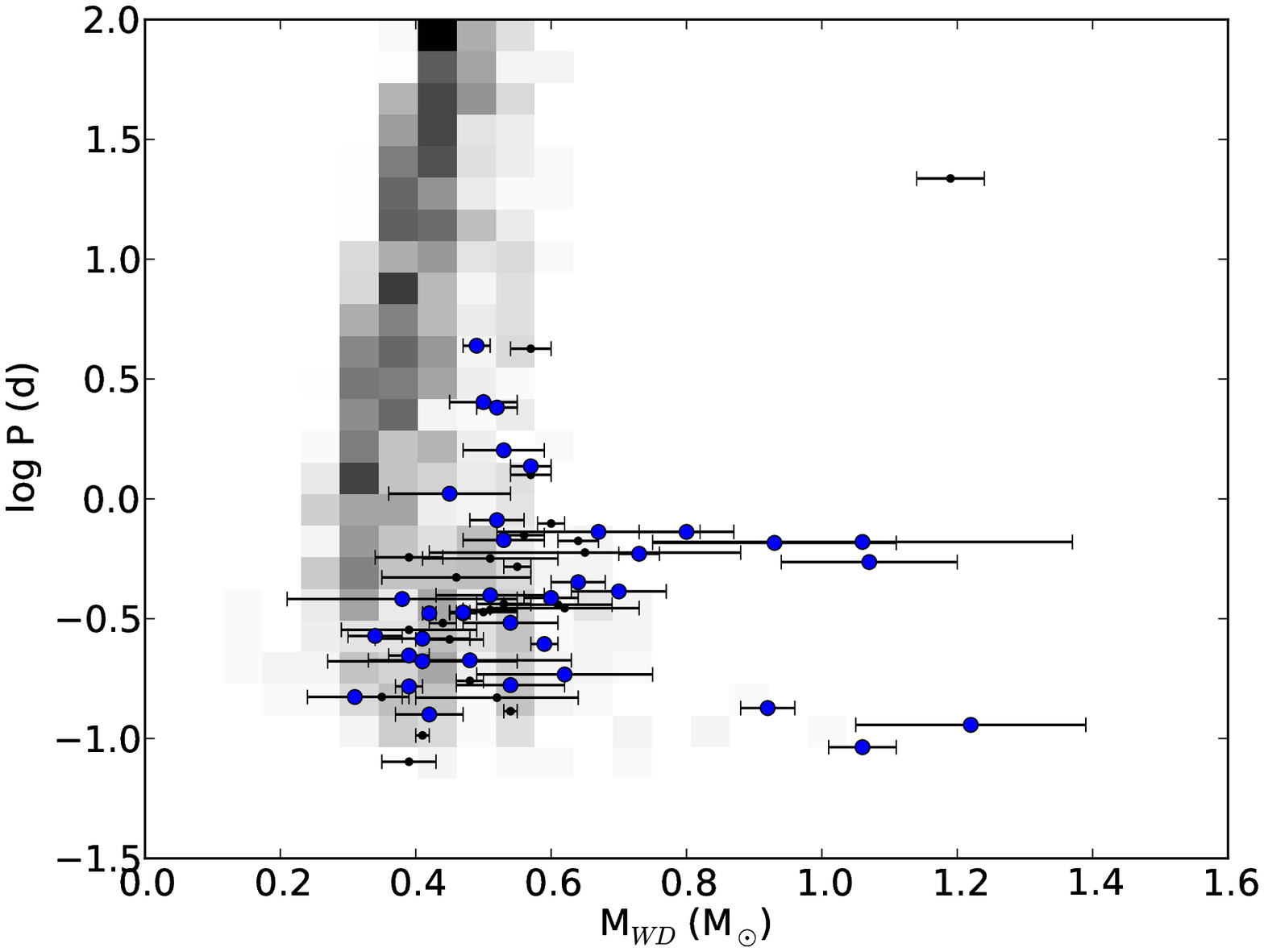} &
	\includegraphics[width=0.5\textwidth]{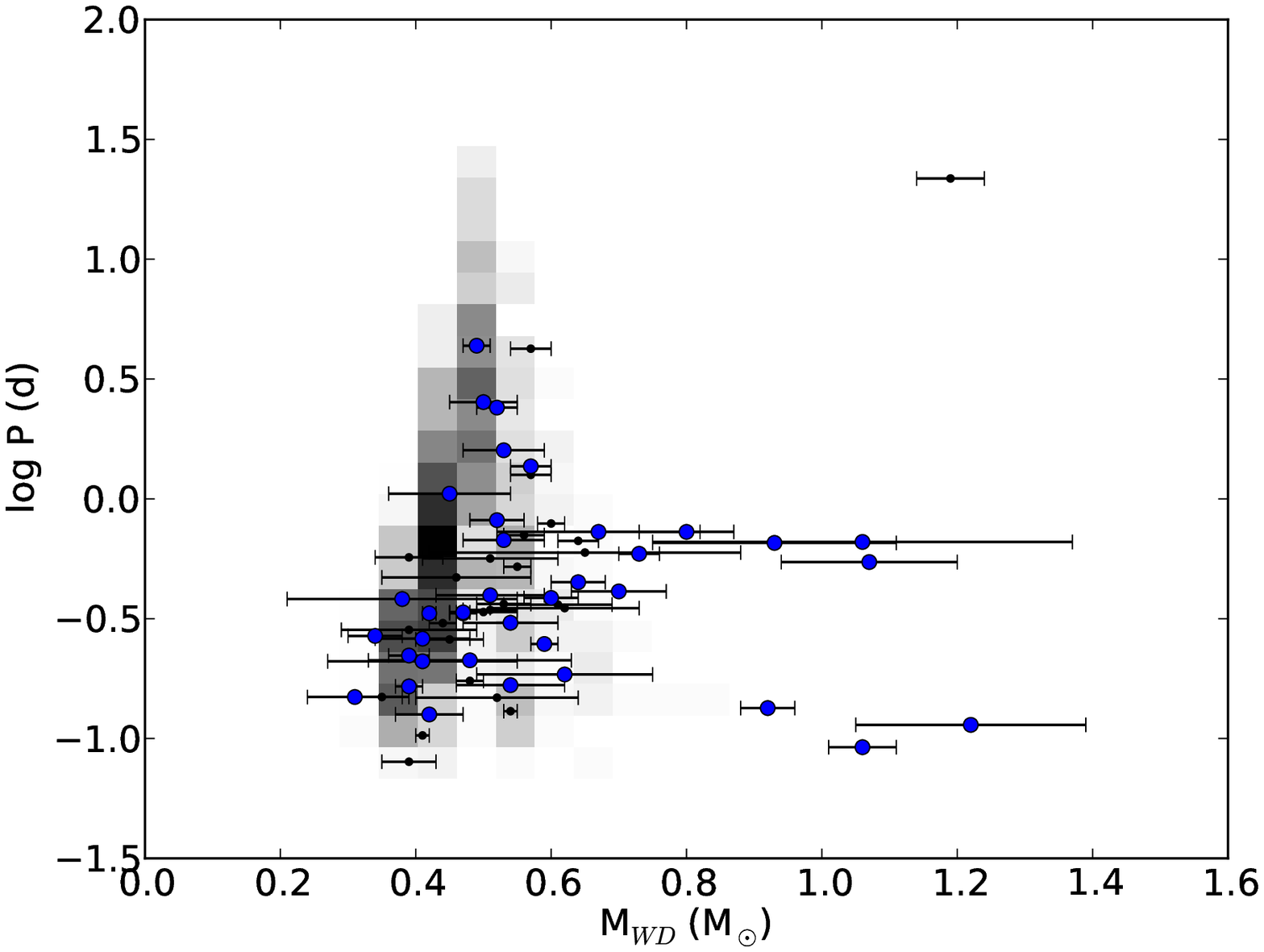} \\
	(c) & (d) \\
	\end{tabular}
    \caption{Visible population of PCEBs in the SDSS as a function of orbital period and WD mass for all models: (a) model $\mga$1, (b) model $\maa$1, (c) model $\mga$2, and (d) model $\maa$2. The intensity of the grey scale corresponds to the density of objects on a linear scale. Overplotted are the observed PCEBs taken from \citet{Zor11b}. Thick points represent systems that are found by the SDSS, and thin points represent previously known PCEBs with accurately measured parameters. The previously known sample of PCEBs is affected by other selection effects than the SDSS sample or the synthetic sample. Note that Ik Peg has been removed from the sample as its MS component  is not an M-dwarf. }
    \label{fig_ch4:pop_P_Mwd_sdss}
    \end{figure*}

To compare our models with the results of \citet{Neb11} and \citet{Zor11}, we place two additional constraints on the visible population of PCEBs in comparison to those described in Sect.\,\ref{sec_ch4:vis}. Following these authors, we only consider WDs that are hotter than 12000K and MS stars of the stellar classification M-type. 
However, there is a discrepancy in the relation between spectral type and stellar mass used in those papers and that of \citet{Kra07} due to the uncertainty in stellar radii of low-mass stars. Where the former \citep[based on][]{Reb07} finds that M-type stars have masses of less than 0.472\Msolar, \citet{Kra07} find that the M-dwarf mass range is extended to 0.59\Msolar.
To do a consistent comparison, we will adopt the relation between spectral type and stellar mass of \citet{Reb07} and the relation between magnitudes and spectral types of \citet{Kra07}. The effect of this discrepancy is further discussed in the sect.\,\ref{sec_ch4:ap_spt_m}. 

Comparing the color-color diagrams of the visible PCEB population (see Fig.\,\ref{fig_ch4:color_color}) with that of the fraction that is visible in the SDSS (see Fig.\,\ref{fig_ch4:color_color_sdss}) shows that the observed population is biased toward late-type secondaries and hot WDs \citep[see][Fig.\,2]{Aug08}. The bias against systems containing early-type secondaries is in accordance with the findings of \citet{Reb10} for the WDMS population from the SDSS.

\citet{Neb11} studied the observed period distribution of observed PCEBs.
They find the distribution follows approximately a log-normal distribution that peaks at about 10.3h and ranges from 1.9h to 4.3d (see points in Fig.\,\ref{fig_ch4:pop_Mms_P_sdss}). They also find that 
the period distribution of the PCEBs found by the SDSS is very comparable to that of previously known PCEBs.
However, \citet{Neb11} point out that the dearth of long-period systems is in contradiction with the results of binary population synthesis studies \citep[see][]{deK93, Wil04, Pol07, Dav10} indicating a low $\alpha$-CE efficiency, if selection effects do not play a role.  
Fig.\,\ref{fig_ch4:pop_Mms_P_sdss} shows that the selection effects does not cause a dearth of PCEBs with long periods in model $\maa$1, $\mga$1, and $\mga$2. Only the results of model $\maa$2 with a reduced $\alpha$-CE efficiency are consistent with the observed period distribution. 

Another observational constraint for our models can come from the relative population sizes. Although the space density of PCEBs is not known very accurately, it has become possible to determine the fraction of PCEBs amongst WDMS systems of all periods. 
From SDSS observations, \citet{Neb11} find that the fraction of PCEBs amongst unresolved WDMS with M-dwarf companions is 27$\pm$2\%.  
Wide WDMS that are blended or fully resolved are not included in their sample. 
To compensate for this effect, we exclude those WDMS systems from our WDMS sample for which the angular size of an object is larger than twice the seeing, where the size of the object is approximated by the orbital separation, and where the distance to the WDMS is given by the Galactic model (see sect.\,\ref{sec_ch4:gal}). 
The seeing limit is varied between the median seeing of SDSS of 1.4$''$ and an upper limit for over 90\% of the SDSS data of 2$''$. 
Furthermore, the SDSS PCEB identification method is based on radial velocity measurements, and, as such dependent on the number of spectra taken, the temporal sampling of the measurements and the accuracy of the radial velocity measurement. \citet{Neb11} find that their identification method is not sensitive to systems with periods of more than a few tens of days, however, the observed period distribution that cuts off at a few days is not dramatically affected by this bias. To account for the long-period bias, we exclude long-period PCEBs from the PCEB sample (but include them in the WDMS sample). 
As the sensitivity of the SDSS PCEB identification method depends on the orbital period of the system \citep[see Fig.\,10 of][]{Neb11}, the limiting period is varied between 10d and 50d. 
The fraction of visible PCEBs amongst unresolved WDMS for all models are consistent with the observed value within a factor of two (see Tbl.\,\ref{tbl_ch4:ratio}). Based on a sample of WDs with near-infrared emission observed with the Hubble Space Telescope, \citet{Far10} also found a ratio of about 25\%.

\begin{table}
\caption{The fraction of visible PCEBs amongst unresolved WDMS for different models of CE evolution. The errors are not statistical errors but come from varying boundaries for the limiting period and seeing. }
\begin{tabular}{lc}
\thickhline
Model $\mga$1 & 0.17-0.23 \\
Model $\maa$1 & 0.27-0.35 \\
Model $\mga$2 & 0.10-0.14 \\
Model $\maa$2 & 0.13-0.15 \\
\\
Observed & 0.27 $\pm$ 0.02 $^1$\\
\thickhline
\hline
\end{tabular}
\label{tbl_ch4:ratio}
\tablefoot{$^1$\citet{Neb11}}
\end{table}

\citet{Zor11} studied the mass dependencies of the orbital period distribution and found that systems containing high-mass secondaries tend to have longer orbital periods. Figure\,\ref{fig_ch4:pop_Mms_P_sdss} shows a similar trend in our models, however due to the period distribution model $\maa$2 reproduces the observations best. 
The relation between WD and secondary mass cannot be used to differentiate between CE theories, because the models show very similar distributions (see Fig.\,\ref{fig_ch4:pop_Mms_Mwd_sdss}), which is contrary to the complete visible PCEB sample (see Fig.\,\ref{fig_ch4:pop_Mms_Mwd}). 
However, the models match well to the observed systems in Fig.\,\ref{fig_ch4:pop_Mms_Mwd_sdss} for WD masses less than about 0.7\Msolar. The models show a lack of PCEBs with massive WD components (see sect.\,\ref{sec_ch4:concl} for a discussion). 
Disregarding WDs with masses more than about 0.8\Msolar, Fig.\,\ref{fig_ch4:pop_P_Mwd_sdss} shows a good match between the observations and the predictions of model $\maa$2 regarding the distribution of orbital period versus WD mass. The discrepancy between the observed period distribution and the synthetic ones from model $\maa$1, model $\mga$1, and model $\mga$2 is mainly found in PCEBs with helium WD components.  
For He-core WDs (i.e. $M_{\rm WD} < 0.5$\Msolar), all models show an increase in orbital period with WD mass. Although the observations are not in contradiction with this, statistical evidence of this relation  in the current observed sample has not been found \citep{Zor11}.
In addition, less than half of the observed systems contains a He WD, which is in contradiction with our models for which at least 70\% of PCEB WDs are helium rich (see Table\,\ref{tbl_ch4:hdcd}).

\begin{table}
\caption{Percentage of helium WDs and carbon/oxygen (including oxygen/neon) WDs in visible PCEBs in the SDSS for different models of CE evolution.}
\begin{tabular}{lcc}
\thickhline
& He WD & CO WD \\ 
Model $\mga$1 & 72 &28\\
Model $\maa$1 & 75&25\\
Model $\mga$2 & 83&17\\
Model $\maa$2 & 68&32 \\
\\
Observed$^1$ &33-46 &54-67\\
\thickhline
\hline
\end{tabular}
\label{tbl_ch4:hdcd}
\tablefoot{$^1$\citet{Zor11}. The percentages in the PCEB population that is found by SDSS are similar to the percentages of the full sample of \citet{Zor11} that is given here. The observed type of the WD is determined by its mass with a limiting mass of 0.5\Msolar~for helium WDs, which is consistent with our models. The range in the observed percentages is caused by a few systems in which the stellar type could not be determined unambiguously.}
\end{table}

\subsection{Variable CE efficiency }
Next, we consider the possibility that CE evolution occurs differently for different types of donor stars or types of instabilities. 
We differentiate between red giant (RG) donors and asymptotic giant branch (AGB) donors and between dynamical (DY) and tidal (TI) instabilities.  
The majority of PCEBs is formed through a dynamical instability initiated by a RG (see Table\,\ref{tbl_ch4:evpath}, RG-DY). 
This evolutionary path is not much affected by processes other than the CE phase, as the evolution is relatively simple: For example the donor stars do not suffer from superwinds as AGB stars. For this path, only the PCEBs from model $\maa$2 with a reduced CE efficiency of $\alpha\lambda=0.25$ are consistent with the observed period distribution and its mass dependencies. Any other CE model produces a high number of PCEBs at periods larger than 10d.

Subsequently, we study CE evolution in the other evolutionary channels with two hypotheses. First, we assume that CE interactions with red giant donors suffer from a low CE efficiency and that those with AGB donors suffer from a high CE efficiency (RG-DY and RG-TI according to model $\maa$2, AGB-DY and AGB-TI according to model $\mga$1 or $\maa$1). However, the PCEB population from this hypothesis does not reproduce the observed period and mass distributions significantly better or worse than model $\maa$2. The percentage of systems containing a He WD improves slightly to about 60\% and 50\% for AGB-DY and AGB-TI according to model $\mga$1 and $\maa$1 respectively.
The second hypothesis is that all systems evolving through dynamical instabilities suffer from low CE efficiency and that systems evolving through a tidal instability do not (RG-DY and AGB-DY according to model $\maa$2 and RG-TI and AGB-TI according to model $\maa$1\footnote{Note that model $\mga$1 is identical to model $\maa$1 for systems evolving through a tidal instability.}). However, also this hypothesis does not lead to a significant improvement (or worsening) in the period and mass distributions compared to model $\maa$2.
The percentage of systems containing a He WD is about 75\%. Concluding, at current we cannot constrain the CE evolution or efficiencies of the evolutionary channels RG-TI, AGB-DY and AGB-TI, although the percentage of systems with helium WDs improves when assuming that the CE efficiency is lower when the CE phase is initiated by a RG star than for an AGB star. 

\begin{table}
\caption{Percentage of visible PCEBs in the SDSS from different evolutionary paths for different models of CE evolution. The last column represents the total number of visible systems for each model in our simulations. RG and AGB represent systems in which the CE phase is initiated by a red giant and a AGB star respectively. DY and TI represent systems that evolve through a dynamical or tidal instability, respectively.} 
\begin{tabular}{lccccc}
\thickhline
 				& RG-DY	& RG-TI	& AGB-DY	& AGB-TI	& Total\\
Model $\mga$1 	& 55	& 18	& 5			& 23		& 1958\\
Model $\maa$1 	& 64	& 11	& 11		& 14		& 2967\\
Model $\mga$2 	& 77	& 7		& 8			& 9			& 1390\\
Model $\maa$2	& 61	& 7		& 24		& 8			& 1142\\
\thickhline
\hline
\end{tabular}
\label{tbl_ch4:evpath}
\end{table}

\section{Discussion and conclusion} 
\label{sec_ch4:concl}

We have studied common-envelope evolution by theoretical modelling of the formation and evolution of post-common-envelope binaries with constraints from observations. We have considered four models of CE evolution that differ in the CE prescription and CE efficiency. The SDSS has played an important role in providing the largest and most homogeneous sample of PCEBs, however, the visible population of PCEBs is still affected by strong selection effects. We presented here the first binary population models that consider the selection effects that are inherent to the population of visible PCEBs. 

We find that although selection effects are important, e.g. for the secondary mass distribution, they do not lead to a dearth of long-period systems as is observed \citep[e.g.,][]{Neb11}. 
Furthermore, we find that the main evolutionary path of visible PCEBs in the SDSS consists of a CE phase caused by a red giant that fills its Roche lobe in a dynamically unstable manner. Most importantly, we find that the efficiency for this channel at which orbital energy can be used to expel the envelope in the CE phase is low - to reproduce the observed period distribution with few systems at 10-100d. Secondary evolutionary paths cannot be constrained at present; low and high CE efficiencies for energy consumption or angular momentum consumption are consistent with observations.

Besides the distribution of orbital periods, the results from the model with the reduced $\alpha$-CE efficiency are consistent with the observed space density, the fraction of PCEBs amongst WDMS, and the WD mass vs. MS mass distribution, however, the fraction of PCEBs containing He WDs is overestimated. When assuming that the CE efficiency is higher when the CE phase is initiated by an AGB star rather than a RG star, the fraction of He WDs in PCEBs is in better agreement with the observations. 
At face value, an overestimation of the fraction of He WDs companions in PCEBs exaggerates the importance of the RG-DY channel, however, the conclusion about the low CE efficiency for RG systems is based on the short periods that are observed for PCEBs with He WD components. 

The fraction of He WDs amongst PCEBs depends on the CE efficiency as shown by Table\,\ref{tbl_ch4:hdcd} \citep[see also][]{deK93, Wil04} and the initial distribution of mass ratios and orbital separations. \citet{Wil04} showed that the effect of the CE efficiency and the initial mass ratios on the He-WD ratio for the \textit{full} PCEB population cannot be distinguished, but that the effect of the initial mass ratio distribution for low CE efficiencies ($\alpha\lambda=0.1$) becomes negligible. 
Furthermore the He WD fraction is affected by the cooling curves of WDs. We have adopted the cooling curves of \citet{Hol06, Kow06, Tre11} that assume a carbon and carbon-oxygen composition of the core. However, He-core WDs for a given stellar mass have a longer cooling time as the specific heat is larger (Althaus, Miller Bertolami, priv. comm).
If we systematically underestimated the brightness of He WDs compared to CO WDs, the synthetic fraction of He WDs in visible PCEBs would be even higher. 

So far, we have studied the CE efficiency ($\alpha$) assuming a constant envelope-structure parameter ($\lambda$). In other words we studied the combination $\alpha\lambda$. While the CE efficiency is not well known, the structure parameter has been calculated by several studies 
\citep{Dew00, Xu10, Lov11}. For low mass stars of $M<3$\Msolar, on average at the onset of the CE phase $\lambda \approx 1.1-1.3$ on the RG and $\lambda \approx 0.5-0.8$ on the AGB \citep[][including internal energy in the envelope binding energy]{Van10}. Therefore, our result of a small value for $\alpha\lambda$ is not due to a small value for $\lambda$; the CE efficiency is low. 

The SDSS has observed six (possibly eight) PCEBs with high WD masses of more than 0.8\Msolar. The number of massive WDs is small, but they represent about 10\% of the observed sample. Although our models do create massive WDs with M-dwarf companions, the relative number to other PCEB systems is not reproduced by our models. If the observed number of these systems increases to a statistical significant amount, it would be interesting to look in more detail in the evolution of these systems, because it is hard to envision how to form a high number of these systems with the current IMF and initial mass ratio distribution . It is particularly interesting in the context of CV-progenitors, as WDs in CVs are on average significantly more massive than single WDs \citep[e.g.][]{War95, Sav11}.

Constraints on CE evolution other than this study have come from reconstruction methods of the evolution of observed binaries. 
From observed PCEBs, \citet{Zor10} deduces a value of $\alpha=0.2-0.3$ for the CE efficiency when including the internal energy of the envelope into the energy balance equation. They find that the internal energy is important for CE evolution when the CE phase is initiated by AGB donors, but the effect is not significant for RG donors. We therefore conclude that our results are consistent with those of \citet{Zor10}. 

From reconstructing the evolution of post-CE binaries (mostly pre-SDSS PCEBs and some SdB+MS binaries), \citet{DeM11} found that the CE efficiency decreases with mass ratio ($q=\frac{M_{\rm a}}{M_{\rm d}}$). The effect, however, has not been observed in the SDSS PCEB population \citep{Zor11}. 

\citet{Por13} reconstructs the formation and evolution of the cataclysmic variable HU Aquarius. The two planets that orbit this CV play an important role in constraining the CE efficiency. \citet{Por13} find a low CE efficiency of $\alpha\lambda = 0.45\pm0.17$ for this system consistent with our conclusion. 

From reconstructing the evolution of double He WDs, \citet{Nel00} deduces two constraints on CE evolution. First, CE evolution  occurs very efficiently (i.e. $\alpha\lambda =2$) in a binary system with a giant donor and a WD companion. The physical interpretation of this is that more energy sources than orbital energy are used to expel the envelope. An example of a possible energy source is the internal energy of the envelope including recombination energy \citep[e.g.][]{Han95, Web08}.
Second, neither the $\alpha$-CE prescription nor stable mass transfer is able to explain the observations for the first phase of mass transfer in the evolution of progenitor systems of double He WDs, 
 and therefore, \citet{Nel00} proposed the $\gamma$-prescription.
Furthermore, \citet{Nel01} showed in a BPS study that the population of Galactic double WDs is well modelled when assuming $\gamma=1.75$ and $\alpha\lambda=2$, whereas the standard $\alpha$-prescription does not. 
Recently, \citet[][see also \citealt{Woo10_Myk}]{Woo12} suggested a new evolutionary model to create double WDs that involves stable, non-conservative mass transfer between a RG and a MS star. The effect on the orbit is a modest widening with a result alike to the $\gamma$-description.

Summarizing, the CE phase is a crucially important phase in the formation and evolution of binaries, however, it is not well understood. 
BPS and evolutionary reconstruction studies have lead to valuable constraints on CE evolution that contribute to the formation of a coherent picture of CE evolution over mass ratios and stellar types involved in the CE phase. In our option, the emerging picture of CE evolution with non-degenerate companions thus far is that:
\begin{itemize}
\item in approximately equal mass binaries that lead to the formation of double WDs, mass transfer leads to a modest widening of the orbit;
\item in binaries with low mass ratios ($q\approx 0.2-0.5$) that lead to the formation of PCEBs, CE evolution leads to a strong contraction of the orbital separation;
\item in binaries with extreme low mass ratios ($q\lesssim 0.2$) the CE phase is caused by a tidal instability rather than a dynamical instability and the CE phase might evolve differently for that reason.  
\end{itemize}

\begin{acknowledgements}
We thank Kars Verbeek, Simo Scaringi, and Kevin Covey for very helpful discussions on colours and magnitudes of main-sequence stars. This work was supported by the Netherlands Research Council NWO (grant VIDI [\# 639.042.813]) and by the Netherlands Research School for Astronomy (NOVA).
\end{acknowledgements}

\begin{appendix}
\section{Population synthesis code SeBa}
\label{sec_ch4:ap_seba}
We present here the most important changes that we made to the population synthesis code SeBa, since \citet{Too12}. 
First, the method of modelling a tidal instability \citep{Dar1879} is updated.
This instability takes place in systems of extreme mass ratios in which there is insufficient orbital angular momentum to keep the stars in synchronous rotation \citep{Hut80}.
The tidal forces that are responsible for the orbital decay are strongly dependent on the ratio of the stellar radius and the distance between the stars \citep{Zah77}. Instead of checking at RLOF, we assume tidal forces are effective if the stellar radius is less than one-fifth of the periastron distance between the stars and that the orbital decay proceeds instantaneously. 

In addition, the winds of hydrogen-poor helium-burning stars are updated. We adopt the formalism of \citet{Hur00} which consists of the maximum of the wind mass-loss of \citet{Rei75} and a Wolf-Rayet-like wind-mass loss. 

Finally, the responses of the radius of helium stars to mass loss are updated. The adiabatic response $\zeta_{\rm ad}$ and the thermal response $\zeta_{\rm eq}$ \citep[see Eq.~A.14~and A.18 in][]{Too12} are used to determine the stability of Roche lobe overflow. For helium MS-stars and helium hertzsprung-gap stars, we assume $\zeta_{\rm ad}=4$. For helium giants, $\zeta_{\rm ad}$ is based on the prescription of \citet{Hje87}. 
For helium hertzsprung-gap stars and helium giants, we assume $\zeta_{\rm th}= -2$. 

\section{The spectral type - mass relation}
\label{sec_ch4:ap_spt_m}

\begin{figure*}
    \centering
    \begin{tabular}{ccc}
	\includegraphics[width=0.3\textwidth]{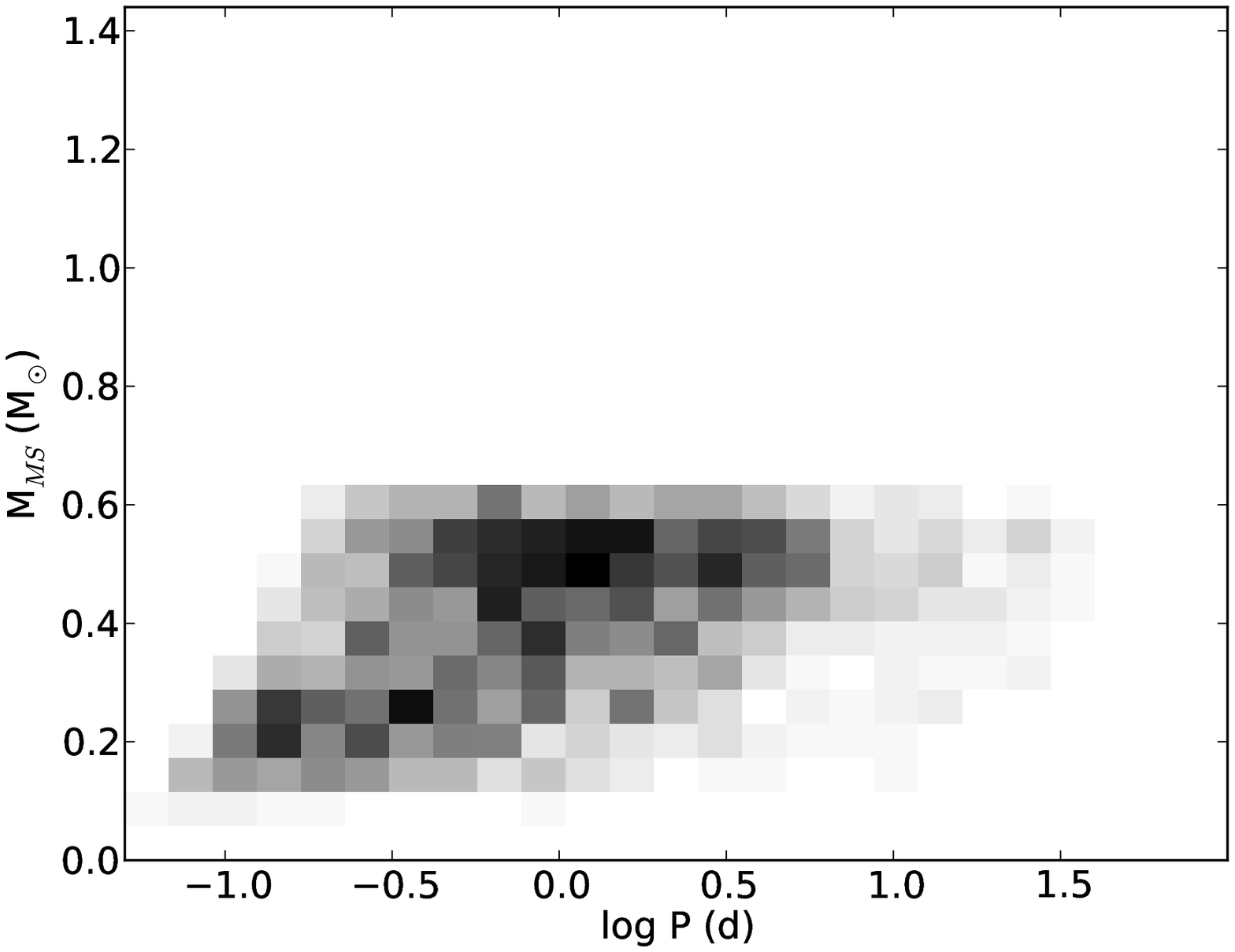} &
	\includegraphics[width=0.3\textwidth]{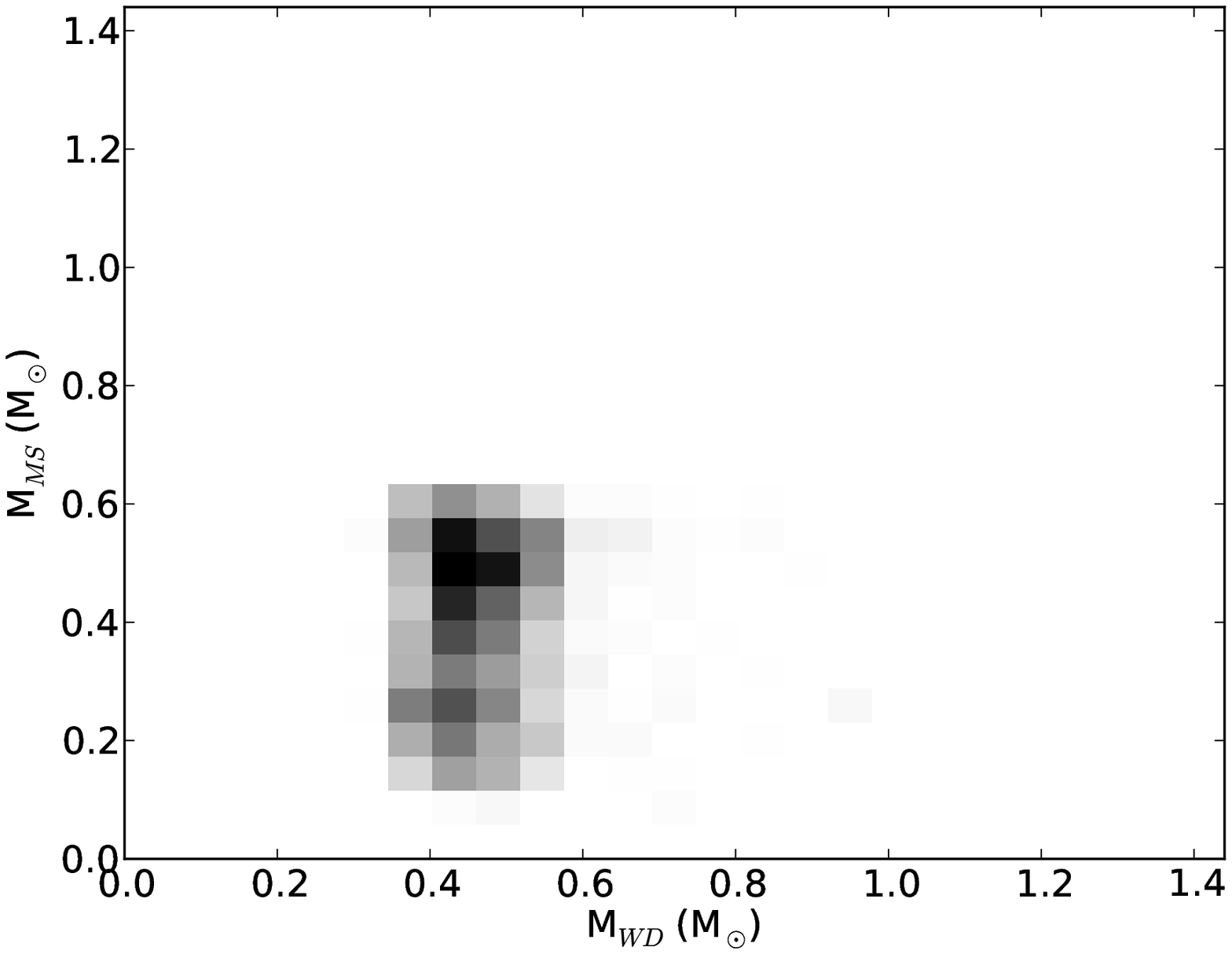} &
	\includegraphics[width=0.3\textwidth]{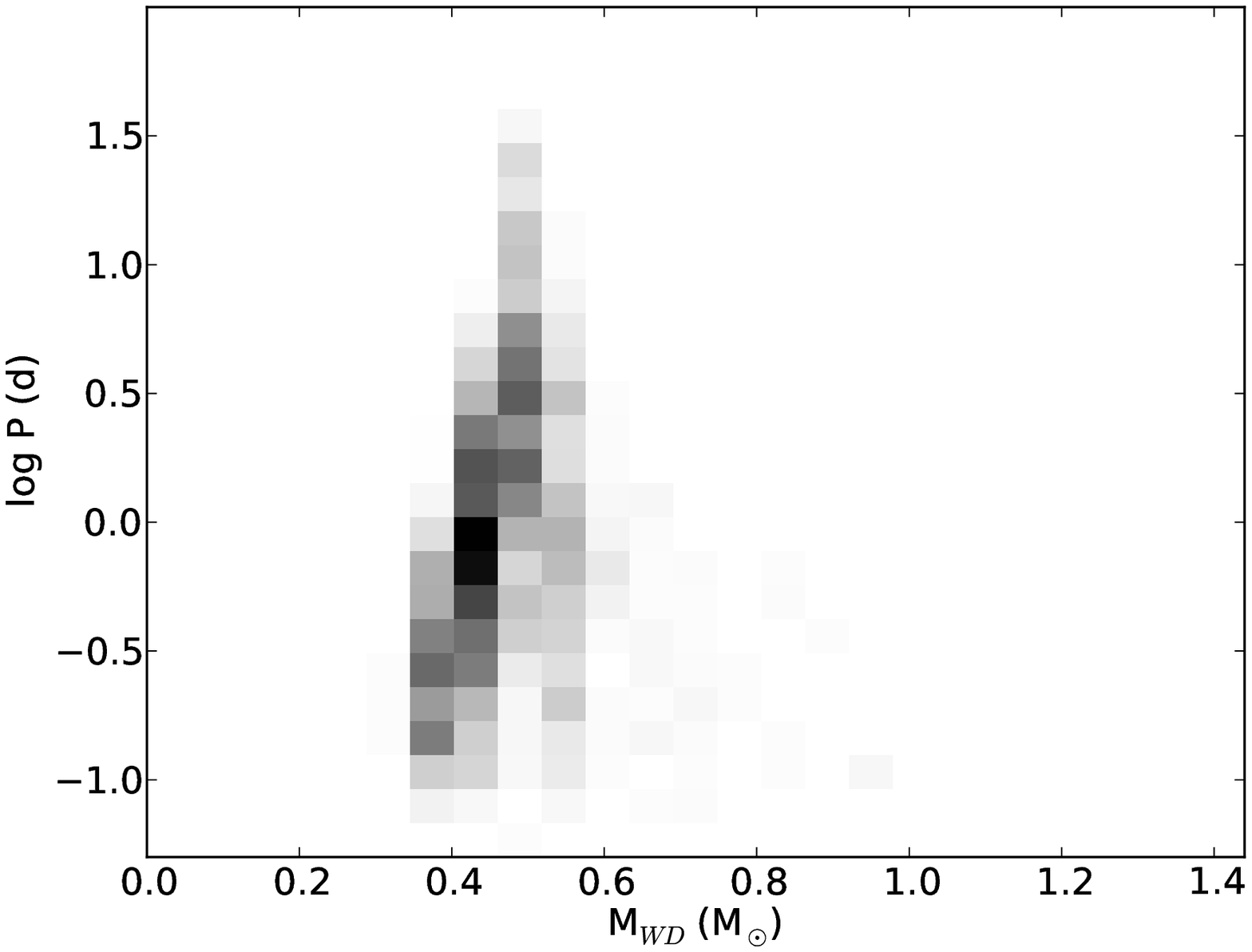} \\
	\end{tabular}
	\caption{Visible population of PCEBs in the SDSS assuming the spectral type-mass relation of \citet{Kra07} for model $\maa2$.}
    \label{fig_ch4:pop_sdss_kraus}
    \end{figure*}

In the last decade, it has become clear that there is a discrepancy between theoretical models of and observationally determined radii and masses of low-mass stars, challenging our understanding of stellar evolution, structure, and atmospheres \citep[see e.g.][]{Hil04,Ber06, Lop07, Boy12}. To exclude these uncertainties from the comparison with the SDSS observations (see sect.\,\ref{sec_ch4:res_sdss}), we have used the relation between spectral type and mass, as determined by \citet{Reb07} and the relation between spectral type and magnitudes, as given by \citet{Kra07}. There is a good agreement between the effective temperatures as a function of spectral type of \citet{Reb07} and \citet{Kra07}. 
For comparison, we show the synthetic populations of visible PCEBs in SDSS in Fig.\,\ref{fig_ch4:pop_sdss_kraus} where we assume the relation between spectral type and mass from \citet{Kra07}. The population is significantly extended to higher secondary masses when compared to those shown in Fig.\,\ref{fig_ch4:pop_Mms_P_sdss}~and~\ref{fig_ch4:pop_Mms_Mwd_sdss}.

\end{appendix}
\bibliographystyle{aa}
\bibliography{bibtex_silvia_toonen_wdms.bib}

\end{document}